\documentclass[3p,twocolumn]{elsarticle}

\usepackage{amssymb}

\usepackage{graphics}

\journal{Astroparticle Physics}

\begin{document}

\begin{frontmatter}



\title{
A Critical Examination on L/E Analysis in the Underground Detectors with a Computer 
Numerical Experiment Part~1
%
%
}

\author[HU]{E.~Konishi\corref{cor1}}
\ead{konish@si.hirosaki-u.ac.jp}
\author[KU]{Y.~Minorikawa}
\author[MoU]{V.I.~Galkin}
\author[MeU]{M.~Ishiwata}
\author[MeU]{I.~Nakamura}
\author[HU]{N.~Takahashi}
\author[ky]{M.~Kato}
\author[SU2]{A.~Misaki}
\address[HU]{
Graduate School of Science and Technology, Hirosaki University, Hirosaki, 036-8561, Japan }    
\address[KU]{
Department of Science, School of Science and Engineering, Kinki University, Higashi-Osaka, 577-8502, Japan }
\address[MoU]{
Department of Physics, Moscow State University, Moscow, 119992, Russia}
\address[MeU]{
Department of Physics, Faculty of Science and Technology,
Meisei University, Tokyo, 191-8506, Japan}
\address[ky]{
Kyowa Interface Science Co.,Ltd., Saitama, 351-0033, Japan }
\address[SU2]{
Innovative Research Organization, Saitama University, Saitama,
 338-8570, Japan}

 \cortext[cor1]{Corresponding author}
\begin{abstract}
 There are many uncertain factors involved in the cosmic ray experiments 
(on atmospheric neutrinos) which seek to confirm the existence of 
neutrino oscillations with definite oscillation parameters. For the 
purpose, it is desirable to select those physical events in which 
uncertain factors are involved to a minimum extent and to examine them in 
a rigorous manner. In the present paper we consider neutrino events due to 
quasi-elastic scattering (QEL) as the most reliable events among various 
candidate events to be analyzed, and have carried out the first step of 
an $L/E$ analysis which aims to confirm the survival probability with a 
Numerical Computer Experiment.  The most important factor in the survival 
probability is $L_\nu/E_\nu$, but this cannot be measured for such 
neutral particles. Instead,  $L_\mu/E_\mu$ is utilized in the $L/E$ 
analysis, where $L_\nu$, $L_\mu$, $E_\nu$ and $E_\mu$ denote the flight 
path lengths of the incident neutrinos, those of the emitted leptons,
the energies of the incident neutrinos and those of the emitted leptons, respectively.
According to our Computer Numerical Experiment, the relation of 
$L_{\nu}/E_{\nu} \approx L_{\mu}/E_{\mu}$  doesn't hold. In 
subsequent papers, we show the results on an $L/E$ analysis with the 
Computer Numerical Experiment based on our results obtained in the 
present paper. 
\end{abstract}


\end{frontmatter}


\section{Introduction}
In the detection of neutrino oscillation of the atmospheric neutrinos by 
underground detectors, one examines the zenith angle distribution  of the 
neutrino events, utilizing the different flight path lengths of 
neutrinos due to the morphology of the Earth, based on the survival 
probability for neutrinos of given flavor.  
For example, for $\nu_\mu \rightarrow \nu_\mu$ , it is given as,
\begin{eqnarray}
\hspace*{-9mm}
P(\nu_{\mu} \rightarrow \nu_{\mu})=
 1- sin^2 2\theta \cdot sin^2
(1.27\Delta m^2 L_{\nu} / E_{\nu} ),  
\end{eqnarray}                                                    
where $\theta$ is the mixing angle between the mass eigenstate and the 
weak eigenstate and $\Delta m^2$ is the difference of the squared mass eigenvalues\cite{Ashie2}. The quantities $L_\nu$ and $E_\nu$ denote 
the flight length of the neutrino from the starting point above the Earth 
to the generation of the neutrino event concerned and the 
neutrino energy, respectively.\\ 

The analysis of the zenith angle distribution of the neutrino events of 
the atmospheric neutrinos should be recognized as a test of an indirect 
proof of the survival probability of a given flavor of neutrino 
oscillation, while the $L/E$ analysis aims at a direct proof of the 
survival probability itself.  In this sense, confirmation of the first 
maximum of oscillation in the survival probability through the $L/E$ 
analysis should be carefully examined in a rigorous manner, 
because the confirmation should be regarded as the ultimate proof of the 
existence of the neutrino oscillation probability.

In the $L/E$ analysis, one tries to observe this survival probability 
directly, through the observation of the first, second and third maxima 
in the oscillations, where $P(\nu_{\mu} \rightarrow \nu_{\mu})$  
have the minimum values. 
There are many candidates of neutrino events for $L/E$ analysis. In 
underground detectors, one can observe different kinds of neutrino 
events, such as Sub-GeV e-like, Multi-GeV e-like, Sub-GeV $\mu$-like,
 Multi GeV $\mu$-like, Multi-ring Sub-GeV $\mu$-like, Multi-Ring Multi-GeV 
$\mu$-like, Upward Stopping Muon events and Upward Through Going Muon 
Events and so on, resulting from the morphology or the nature of the 
physical interaction - elastic scattering off an electron, quasi elastic 
scattering off a nucleon, single meson production, deep inelastic 
scattering, coherent pion production and so on\cite{Ashie2}.

Considering that there are many uncertain factors in the study of 
neutrino oscillations with the use of the cosmic ray beam (atmospheric 
neutrinos), one should carry out the neutrino oscillation study with the 
neutrino events of least ambiguity among the possible candidates to be 
analyzed. Thus, it is reasonable to select those quasi-elastic scattering 
(QEL) events classified as {\it Fully Contained Events}, where the 
generation points of the interaction as well as their termination points 
remain inside the detector. 
In addition to its simple shape, from the nature of the physical 
interaction and also from the viewpoint of the morphology, the directions 
of the emitted leptons and their energy can be determined more reliably 
than in the case of any other candidates.

\section{The characteristics of the computer numerical experiment
}
In the present paper, we carry out the first step of $L/E$ analysis of the 
neutrino events which are only due to quasi elastic scattering (QEL), 
such as $\nu + N \rightarrow N + lepton$, and are confined fully 
inside the 
detectors ({\it Fully Contained Events}) with a Computer Numerical 
Experiment. In addition to their high quality, their 
frequency is also highest among all possible events and, consequently, 
statistics of the events for the purpose of extracting definite 
conclusions about neutrino oscillation are not a worry. 
One does not have to rely on other experimental data, based on 
insufficient information, to increase statistics.
In our Computer Numerical Experiment, we "construct" virtually the 
underground detector in the computer and sample randomly a neutrino 
energy from the energy spectrum of incident neutrinos on the Earth,
 follow the neutrino events giving rise to QEL, considering the interior 
structure of the Earth, "measure" both the directions and the energies of 
the emitted leptons (electrons and muons) in the QEL events and finally 
confirm whether these events are really confined inside the detector or 
escape from it by following the emitted leptons in a stochastic manner. 
In other words, we reproduce the neutrino events due to 
atmospheric neutrino inside the computer. In this sense we obtain pseudo 
experimental data, in contrast to the real experimental data in the 
underground detectors. As we adopt the same initial or boundary 
conditions which is applied to the real underground detectors, such as 
detector geometry\cite{Ashie2}, live days for the experiment\cite{Ashie2}, 
incident neutrino energy spectrum\cite{Honda} and the interior 
structure of the Earth\cite{Gandhi}, we can 
say that our pseudo experimental data are directly comparable with the 
corresponding real experimental data, allowing for the uncertainties 
occurring in the real experiments.  

Due to the characteristics of the Computer Numerical Experiment, the 
pseudo experimental data thus obtained have no "experimental errors". 
However, even if the phenomena concerned may be proved to be 
realized with the Computer Numerical Experiment, these phenomena are not 
always detected in the real experiments. They may be detected in some 
case, maybe not in another, depending on the constraints of the 
real detectors concerned. Thus, careful and detailed examination of the 
analysis of neutrino events due to QEL which are fully contained in the 
detector are essentially important for the sake of the examination of 
the most reliable $L/E$ analysis. 

Verification of the existence of neutrino oscillations through $L/E$ 
analysis, if they exist, means confirming the existence of the maximum 
oscillation where $P(\nu_{\mu} \rightarrow \nu_{\mu})$ have minimum 
values. It should be noticed that the arguments $L_\nu$ and $E_\nu$  
contained in the survival probability cannot be measured for neutral 
particles and instead, $L_\mu$ and $E_\mu$ are utilized as estimators of 
$L_\nu$ and $E_\nu$, where $L_\nu$ is the flight path length of the 
produced lepton and $E_\nu$ is its energy.

Therefore, in the study on the detection of the survival probability it 
is implicitly assumed that the following equation holds.
\begin{eqnarray}
L_{\nu}/E_{\nu} \approx L_{\mu}/E_{\mu}
\end{eqnarray}                                                    
 								
In the present paper, we focus our attention on the examination of the 
validity of Eq.(2) through the analysis of the 
{\it Fully Contained Events} among QEL events with the Computer 
Numerical Experiment.

\begin{figure}
\begin{center}
\vspace{-1.0cm}
\hspace*{-0.8cm}
\rotatebox{90}{%
\resizebox{0.35\textwidth}{!}{\includegraphics{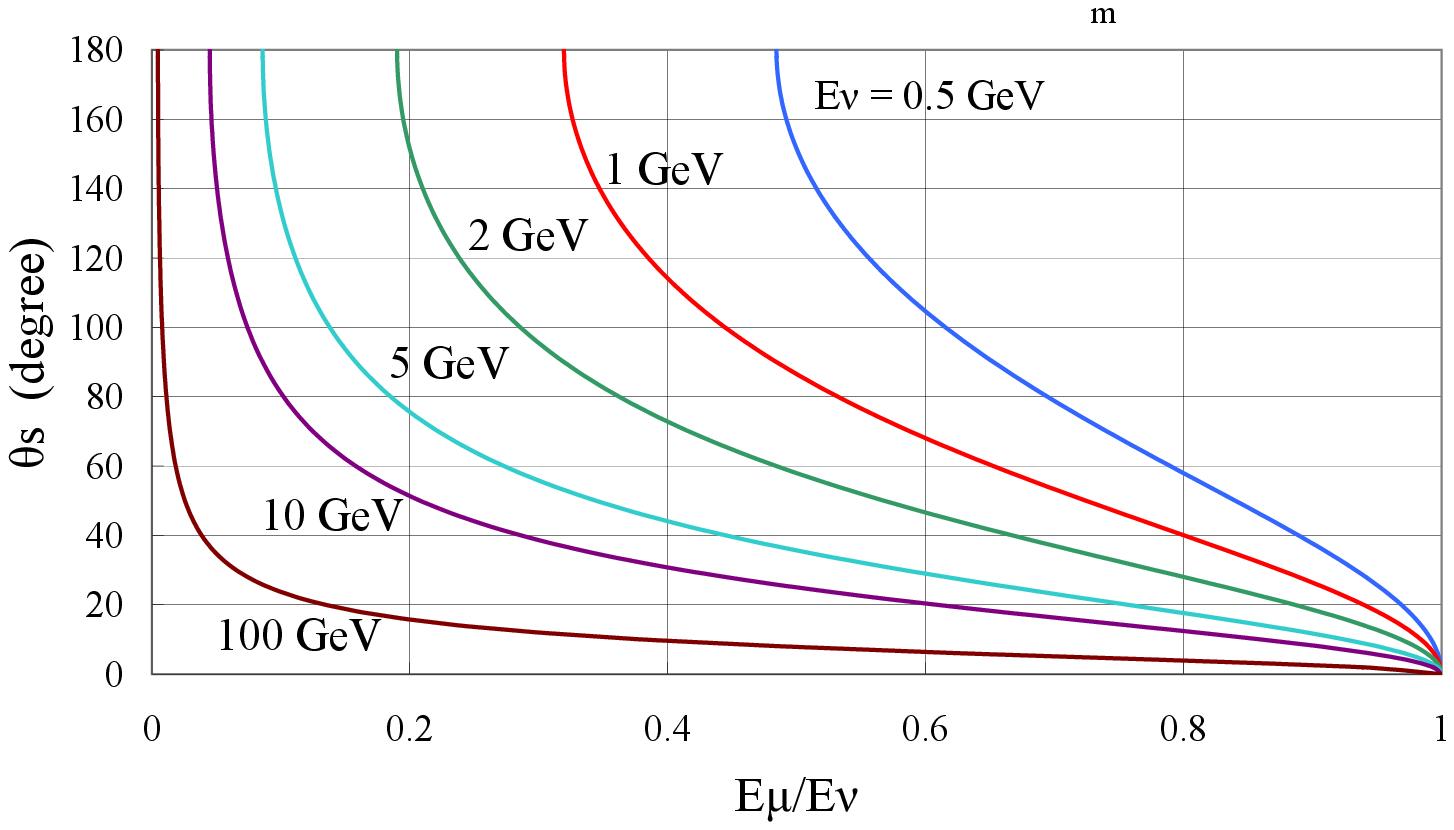}}}

\vspace{-1cm}
\caption{The relation between the energy of the muon and its 
production angle for different incident muon neutrino energies,
 0.5, 1, 2, 5, 10 and 100~GeV.}
\label{figH001}
\vspace{-0.3cm}
\hspace*{-0.8cm}
\rotatebox{90}{%
\resizebox{0.45\textwidth}{!}{\includegraphics{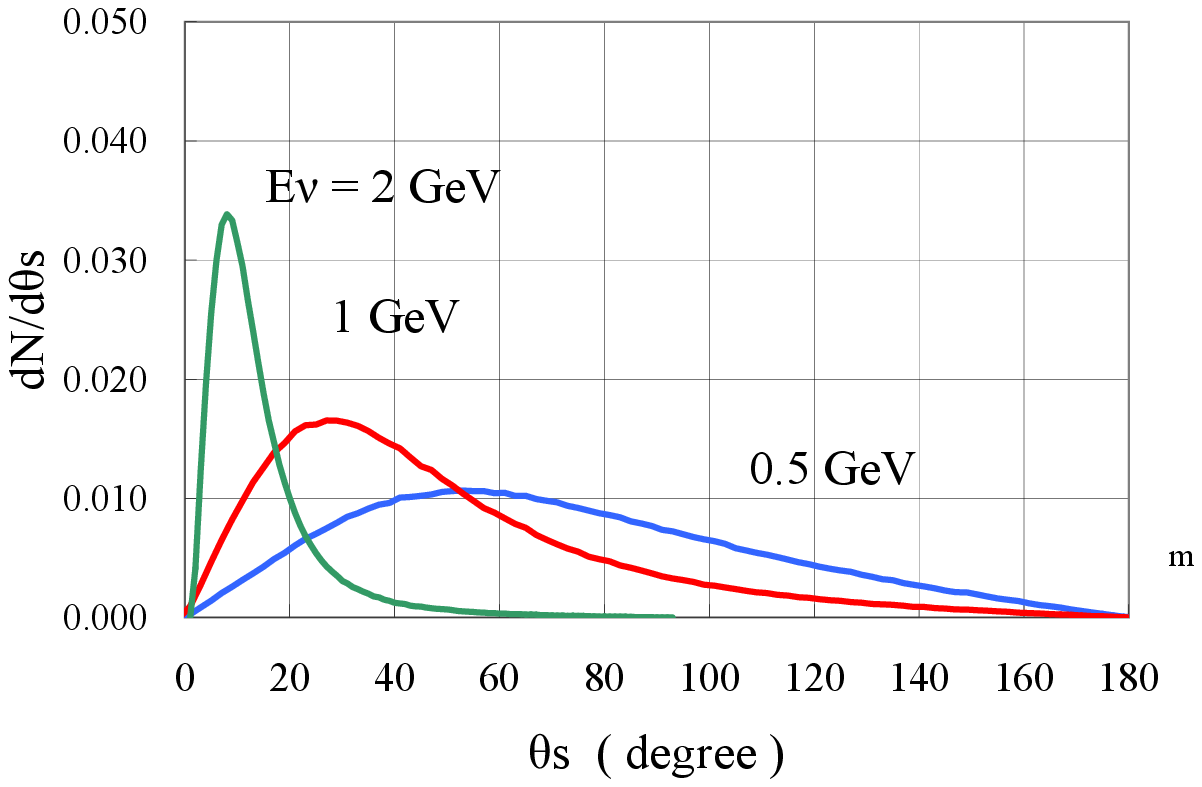}}}
\vspace{-1.5cm}
\caption{The distribution functions of the scattering. The angles of 
the muon for muon-neutrino with incident energies, 0.5 , 1.0 and 
2~GeV. Each curve is obtained by the Monte Carlo 
method (one million sampling per curve). }
\label{figH002}
\end{center}
\end{figure} 

\section{Neutrino events due to quasi-elastic scattering which are fully 
contained in the detector
}
\subsection{
Differential cross section for  quasi-elastic scattering and 
spreads of the scattering angles
}
Here we obtain the distribution functions for the scattering angles of 
the produced leptons. We examine the following quasi elastic scattering 
(QEL) events:
   \begin{eqnarray}
         \nu_e + n \longrightarrow p + e^-  \nonumber\\
         \nu_{\mu} + n \longrightarrow p + \mu^- \nonumber\\
         \bar{\nu}_e + p \longrightarrow n + e^+ \\
         \bar{\nu}_{\mu}+ p \longrightarrow n + \mu^+ \nonumber
         ,\label{eqn:1}
   \end{eqnarray}

The differential cross section for QEL is given as follows\cite{r4}.\\
    \begin{eqnarray}
         \frac{ {\rm d}\sigma_{\ell(\bar{\ell})}(E_{\nu(\bar{\nu})}) }{{\rm d}Q^2} = \nonumber\\
         \frac{G_F^2{\rm cos}^2 \theta_C}{8\pi E_{\nu(\bar{\nu})}^2}
         \Biggl\{ A(Q^2) \pm B(Q^2) \biggl[ \frac{s-u}{M^2} \biggr]
          \nonumber \\ 
+ C(Q^2) \biggl[ \frac{s-u}{M^2} \biggr]^2 \Biggr\},
         \label{eqn:2}
    \end{eqnarray}

\noindent where
    \begin{eqnarray*}
      A(Q^2) &=& \frac{Q^2}{4}\Biggl[ f^2_1\biggl( \frac{Q^2}{M^2}-4 \biggr)+ f_1f_2 \frac{4Q^2}{M^2} \\
 &&+  f_2^2\biggl( \frac{Q^2}{M^2} -\frac{Q^4}{4M^4} \biggr) + g_1^2\biggl( 4+\frac{Q^2}{M^2} \biggl) \Biggr], \\
      B(Q^2) &=& (f_1+f_2)g_1Q^2, \\
      C(Q^2) &=& \frac{M^2}{4}\biggl( f^2_1+f^2_2\frac{Q^2}{4M^2}+g_1^2 \biggr).
    \end{eqnarray*}
\\
\noindent The signs $+$ and $-$ refer to $\nu_{\mu(e)}$ and $\bar{\nu}_{\mu(e)}$ for charged current (c.c.) interactions, respectively.  The $Q^2$ denotes the four momentum transfer between the incident neutrino and the charged lepton. Details of other symbols are given in \cite{r4}.

The relation among $Q^2$, $E_{\nu(\bar{\nu})}$, energy of the incident 
neutrino, $E_{\ell(\bar{\ell})}$, energy of the emitted charged lepton 
(muon orelectron or their anti-particles) and $\theta_{\rm s}$, 
scattering angle of the emitted lepton, is given as
\begin{equation}
         Q^2 = 2E_{\nu(\bar{\nu})}E_{\ell(\bar{\ell})}
(1-{\rm cos}\theta_{\rm s}).
                  \label{eqn:3}
\end{equation}

\noindent Also, energy of the emitted lepton is given by
\begin{equation}
         E_{\ell(\bar{\ell})} = E_{\nu(\bar{\nu})} - \frac{Q^2}{2M}.
\label{eqn:4}
\end{equation}
 
\begin{table*}
\vspace{-2mm}
\caption{\label{tab:table1} The average values $<\theta_{\rm s}>$ for 
scattering angle of the emitted charged leptons and their standard 
deviations $\sigma_{\rm s}$ for various primary neutrino energies 
$E_{\nu(\bar{\nu})}$.}
\vspace{2mm}
\begin{center}
\begin{tabular}{|c|c|c|c|c|c|}
\hline
&&&&&\\
$E_{\nu(\bar{\nu})}$ (GeV)&angle&$\nu_{\mu(\bar{\mu})}$&
$\bar{\nu}_{\mu(\bar{\mu})}$&$\nu_e$&$\bar{\nu_e}$ \\
&(degree)&&&&\\
\hline
0.2&$<\theta_\mathrm{s}>$&~~ 89.86 ~~&~~ 67.29 ~~&~~ 89.74 ~~&~~ 67.47 ~~\\
\cline{2-6}
   & $\sigma_\mathrm{s}$ & 38.63 & 36.39 & 38.65 & 36.45 \\
\hline
0.5&$<\theta_\mathrm{s}>$& 72.17 & 50.71 & 72.12 & 50.78 \\
\cline{2-6}
   & $\sigma_\mathrm{s}$ & 37.08 & 32.79 & 37.08 & 32.82 \\
\hline
1  &$<\theta_\mathrm{s}>$& 48.44 & 36.00 & 48.42 & 36.01 \\
\cline{2-6}
   & $\sigma_\mathrm{s}$ & 32.07 & 27.05 & 32.06 & 27.05 \\
\hline
2  &$<\theta_\mathrm{s}>$& 25.84 & 20.20 & 25.84 & 20.20 \\
\cline{2-6}
   & $\sigma_\mathrm{s}$ & 21.40 & 17.04 & 21.40 & 17.04 \\
\hline
5  &$<\theta_\mathrm{s}>$&  8.84 &  7.87 &  8.84 &  7.87 \\
\cline{2-6}
   & $\sigma_\mathrm{s}$ &  8.01 &  7.33 &  8.01 &  7.33 \\
\hline
10 &$<\theta_\mathrm{s}>$&  4.14 &  3.82 &  4.14 &  3.82 \\
\cline{2-6}
   & $\sigma_\mathrm{s}$ &  3.71 &  3.22 &  3.71 &  3.22 \\
\hline
100&$<\theta_\mathrm{s}>$&  0.38 &  0.39 &  0.38 &  0.39 \\
\cline{2-6}
   & $\sigma_\mathrm{s}$ &  0.23 &  0.24 &  0.23 &  0.24 \\
\hline
\end{tabular}
\end{center}
\label{tab:1}
\vspace{-2mm}
\end{table*}
Now, let us examine the magnitude of the scattering angle of the emitted 
lepton in a quantitative way, as this plays a decisive role in 
determining the accuracy of the direction of the incident neutrino, 
which is directly related to the reliability of the zenith angle 
distribution of {\it Fully Contained muon (electron) Events} 
in the underground detectors. Using a Monte Carlo method and equations 
(4) to (6), we obtain the distribution function for the scattering angle 
of the emitted leptons and the related quantities. The procedure 
for determining the scattering angle for a given energy of the incident 
neutrino is described in Appendix A.

   Figure~\ref{figH001} shows this relation for muons, in which we can 
easily understand that the scattering angle $\theta_{\rm s}$ of the 
emitted lepton (muons here) is greatly influenced not only by the primary 
neutrino energy but by the emitted lepton's fractional energy.
 The effect of scattering angles of the emitted muons (leptons) on the 
determination of the direction of incident neutrinos therefore cannot be 
neglected. 
For a quantitative examination of the scattering angle, we construct the 
distribution of $\theta_{\rm s}$ of the emitted lepton. 
   Figure~\ref{figH002} gives the distribution function for 
$\theta_{\rm s}$  of the muon produced in QEL interactions. 
It can be seen that the muons produced from lower energy neutrinos are 
scattered over wider angles and that a considerable part of them are 
scattered even in backward directions. Similar results are obtained for 
muon anti-neutrinos, electron neutrinos and anti-electron neutrinos.

Also, in a similar manner, we obtain not only the distribution function 
for the scattering angle of the charged leptons, but also their average 
values $<\theta_{\rm s}>$ and their standard deviations $\sigma_{\rm s}$. 
Table~1 shows these quantities for muon neutrinos, anti-muon neutrinos, 
electron neutrinos and anti-electron neutrinos.  

   Table~1 shows these quantities for muon neutrinos, muon anti-neutrinos, electron neutrinos and electron anti-neutrinos. 
From Table 1, Figure~\ref{figH001} and Figure~\ref{figH002}, 
it is clear that for the purpose of establishing the 
incident neutrino's direction the scattering angle of the emitted lepton 
cannot be neglected, especially taking into account the fact that the 
frequency of the neutrino events with smaller energies is far larger than 
that of the neutrino events with larger energies due to the steepness of 
the neutrino energy spectrum. 
Also, it is clear from Table~1 that there are almost no differences 
between average scattering angles, and their standard deviations, for 
muons (anti-muons) and electrons (anti-electrons). 
The difference of the rest masses between electrons and muons can be 
completely neglected at the energies under consideration. This 
fact plays an important role in the discrimination of muons from 
electron. We shall discuss this matter in Part~2 of our paper
\footnote{The discrimination of an electron with a given energy from a 
muon with the same energy is only possible by distinguishing the shape 
of the photon distributions due to the electrons from that of the 
corresponding muons. With regard to this problem, see a subsequent 
paper (Part~2).}.

\subsection{
The relation between the direction of the emitted lepton and that of the 
incident neutrinos
}
In the previous subsection, we point out the non-negligible effect of 
finite scattering angles of the emitted lepton on the estimation of the 
direction of the incident neutrinos. Here, we emphasize that the 
deviation of the directions of emitted leptons from those of the incident 
neutrinos are enhanced by the effect of the azimuthal angle of the 
scattering. 
We examine the effect of the azimuthal angle, $\phi$, and the scattering angle $\theta_{\rm s}$ of an emitted lepton over its zenith angles,
$\theta_{\mu({\bar\mu})}$, for given zenith angles of the incident neutrinos, $\theta_{\nu({\bar\nu})}$ in QEL.

\begin{figure}
\begin{center}
\resizebox{0.5\textwidth}{!}{%
  \includegraphics{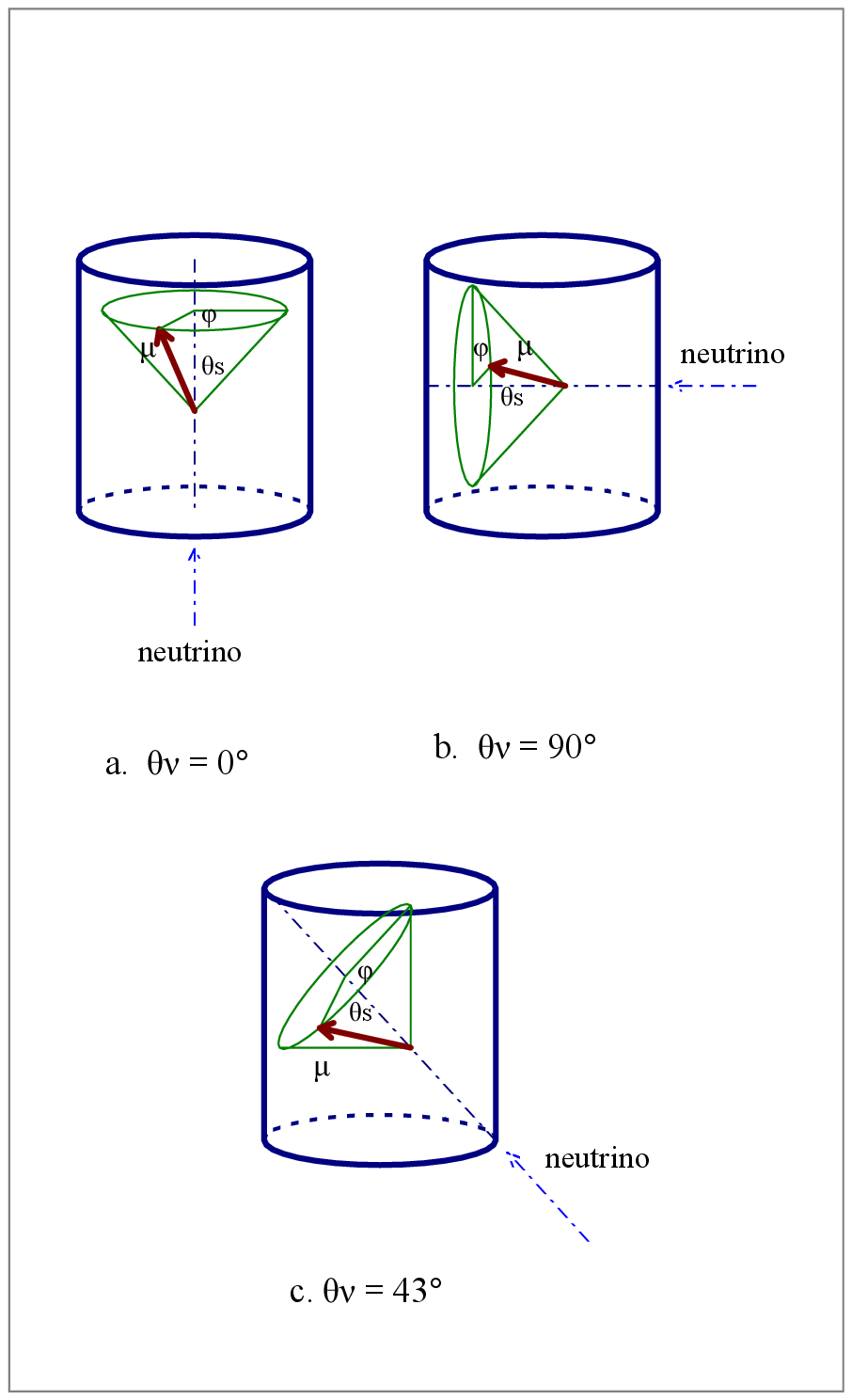}
  }
\vspace{-1.0cm}
\caption{A schematic view of the zenith angles of the charged
 muon for different zenith angles of the incident neutrino, focusing on
  their azimuthal angles.}
\label{figH003}
\end{center}
\end{figure} 

 For three typical cases (vertical, horizontal and diagonal), 
Figure~\ref{figH003} gives a schematic representation of 
the relationship between the zenith angle of the incident neutrino
,$\theta_{\nu({\bar\nu})}$, and the pair of scattering angles,
$(\theta_{\rm s}, \phi)$ of the emitted muon. 
The zenith angle of the emitted muon is derived from 
$\theta_{\nu({\bar\nu})}$ and $(\theta_{\rm s}, \phi)$
  by (A.6) as shown in Appendix A.

 From Figure~\ref{figH003}-a, it can been seen that the zenith angle
 $\theta_{\mu({\bar\mu})}$ of the emitted lepton is not influenced by 
its value of $\phi$ for vertically incident neutrinos 
($\theta_{\nu({\bar\nu})}$ = 0‹), as expected.
 From Figure~\ref{figH003}-b however, it is 
obvious that the influence of $\phi$ of the emitted leptons on their own 
zenith angle is the strongest in the case of horizontal incidence of the 
neutrino ($\theta_{\nu({\bar\nu})}$ = 90‹). In that case one half of the 
emitted leptons are recognized as upward going, while the other half is 
classified as downward going. 
In Figure~\ref{figH003}-c we give the diagonal case 
($\theta_{\nu({\bar\nu})}$ = 43‹). It shows the intermediate situation 
between the vertical and the horizontal. 
In the following, we examine the cases for vertical, horizontal and 
diagonal incidence of neutrinos with different energies,
$E_{\nu({\bar\nu})}$ =~0.5~GeV, 1~GeV and 5~GeV as typical cases.

\begin{figure*}
\vspace{-1.0cm}
\hspace{2.5cm}(a)
\hspace{5.5cm}(b)
\hspace{5.5cm}(c)
\vspace{-0.3cm}
\begin{center}
\resizebox{\textwidth}{!}{%
  \includegraphics{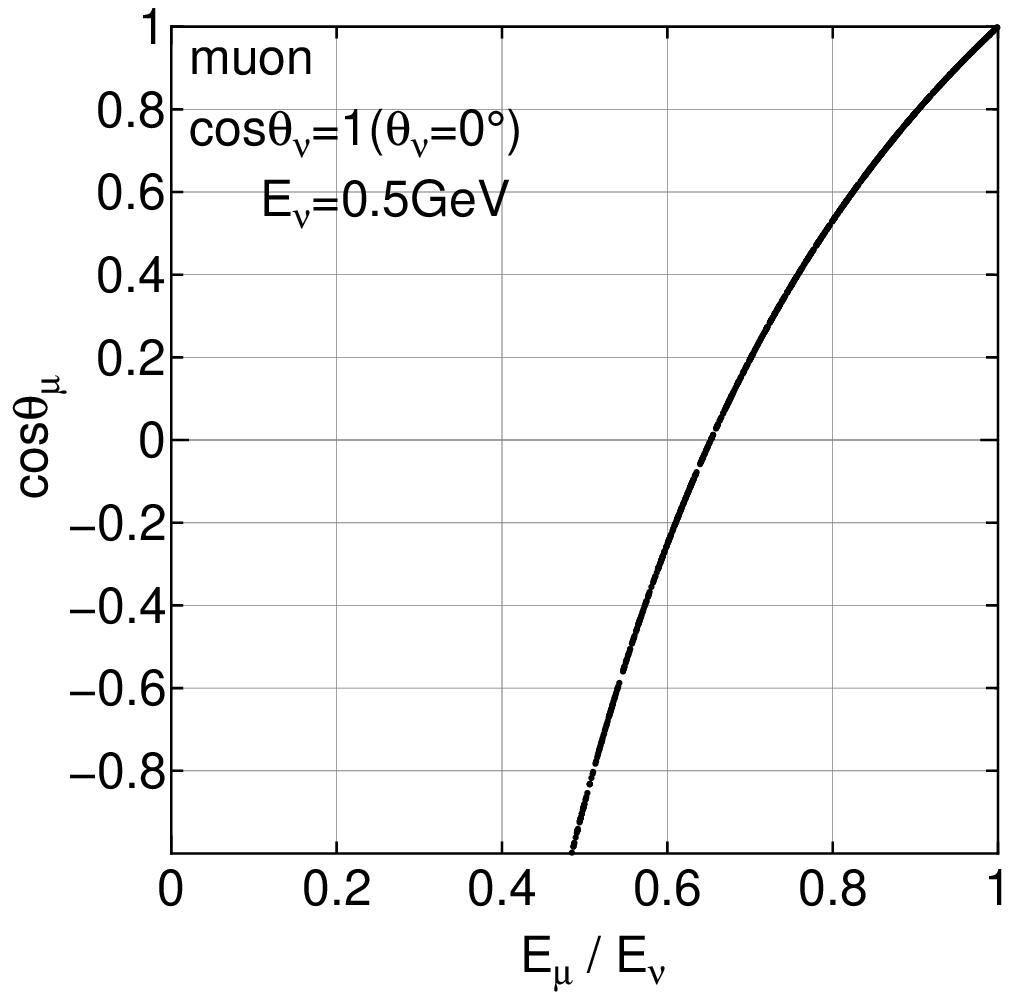}\hspace{1cm}
  \includegraphics{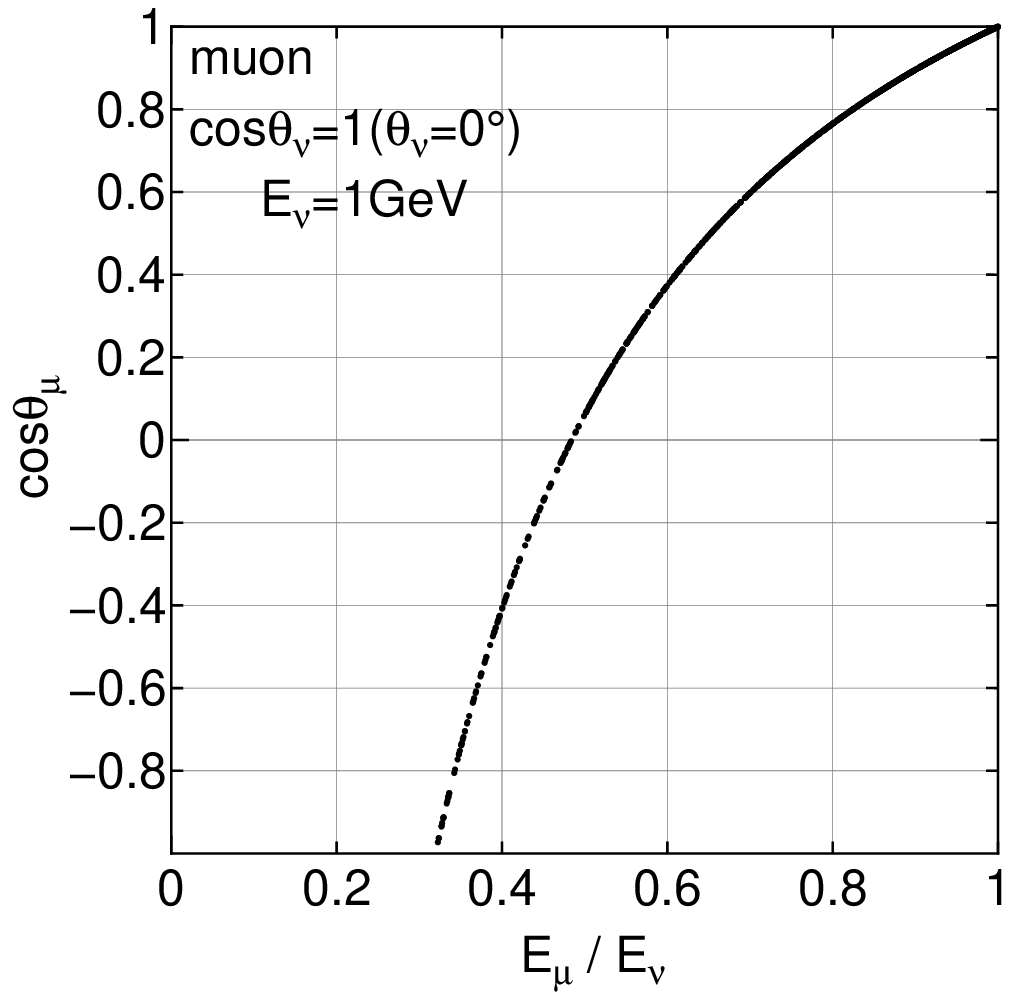}\hspace{1cm}
  \includegraphics{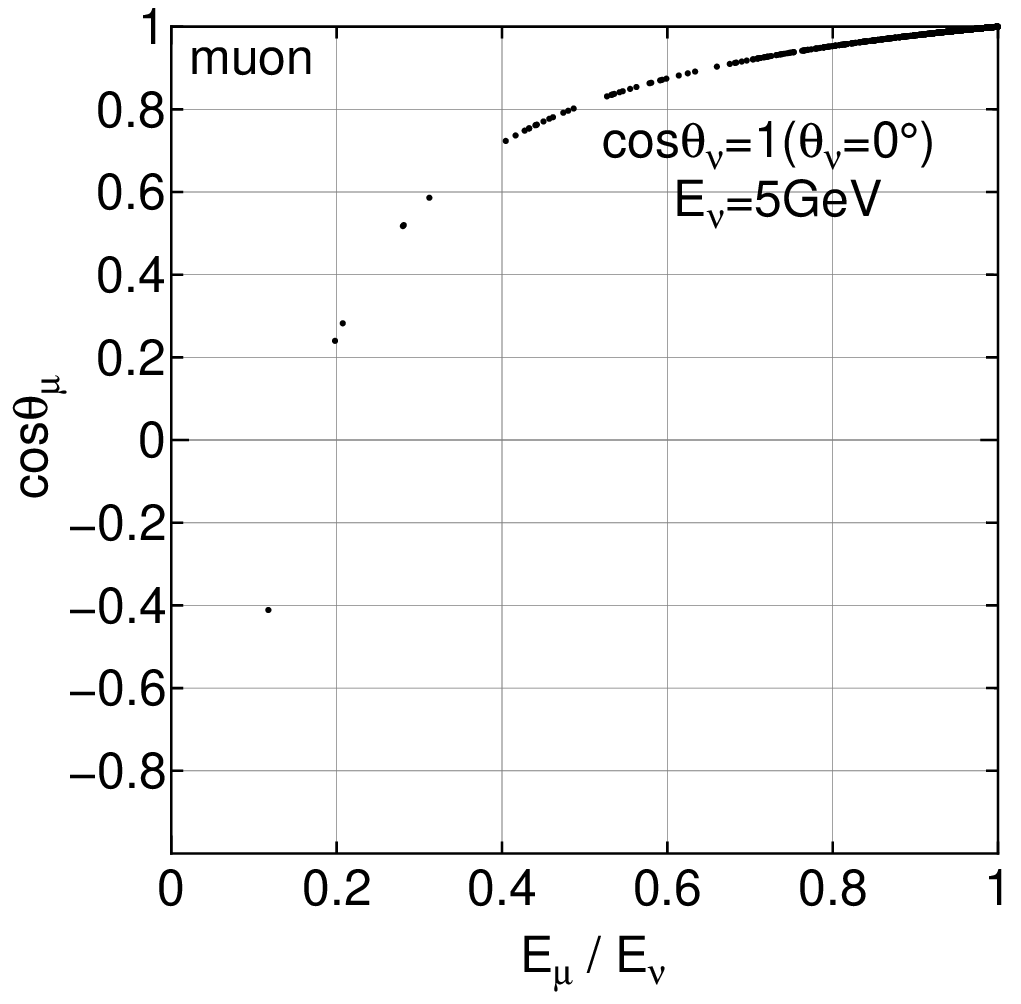}
}
\caption{
\label{figH004}
The scatter plots of the fractional energies of the produced muons 
and their zenith angles for vertically incident muon neutrinos 
with 0.5~GeV, 1~GeV and 5~GeV, respectively.
 The sampling number is 1000 in each case.
}
\end{center}
\vspace{0.5cm}
\hspace{2.5cm}(a)
\hspace{5.5cm}(b)
\hspace{5.5cm}(c)
\vspace{-0.3cm}
\begin{center}
\resizebox{\textwidth}{!}{%
  \includegraphics{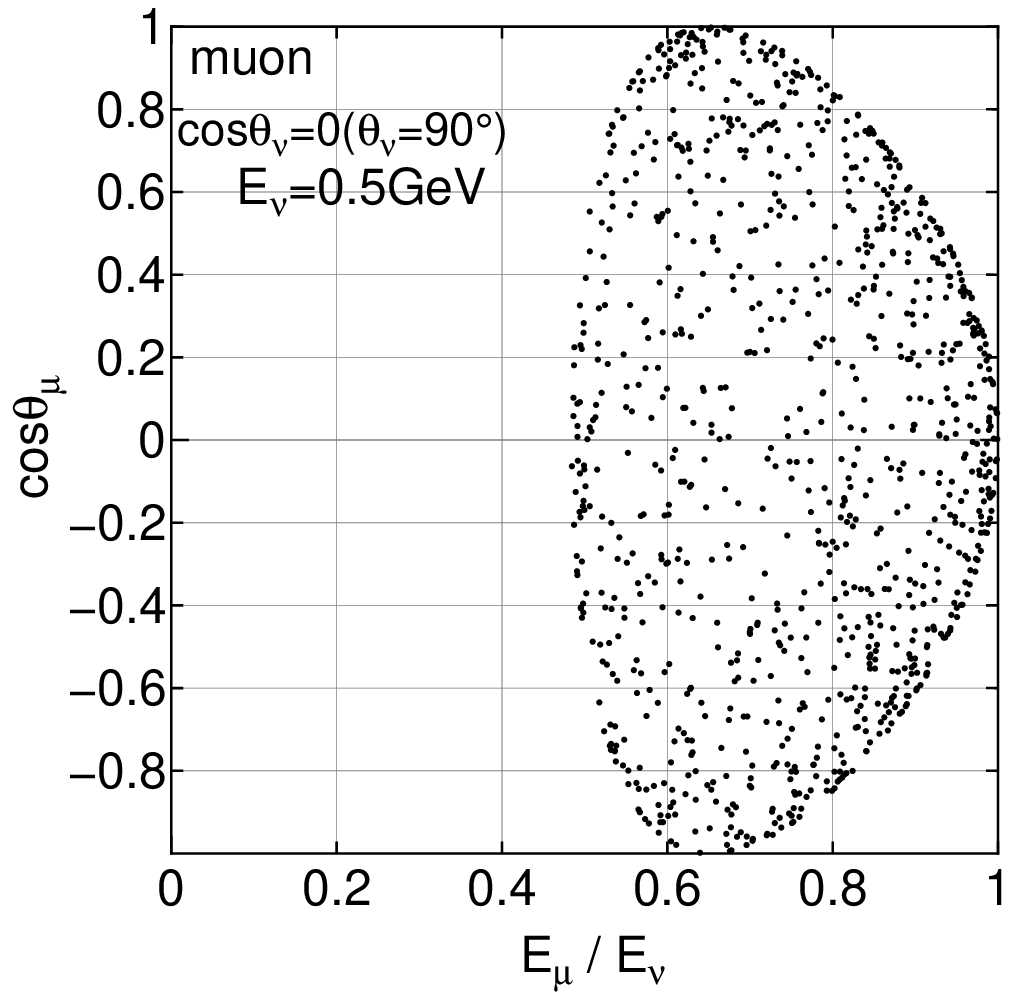}\hspace{1cm}
  \includegraphics{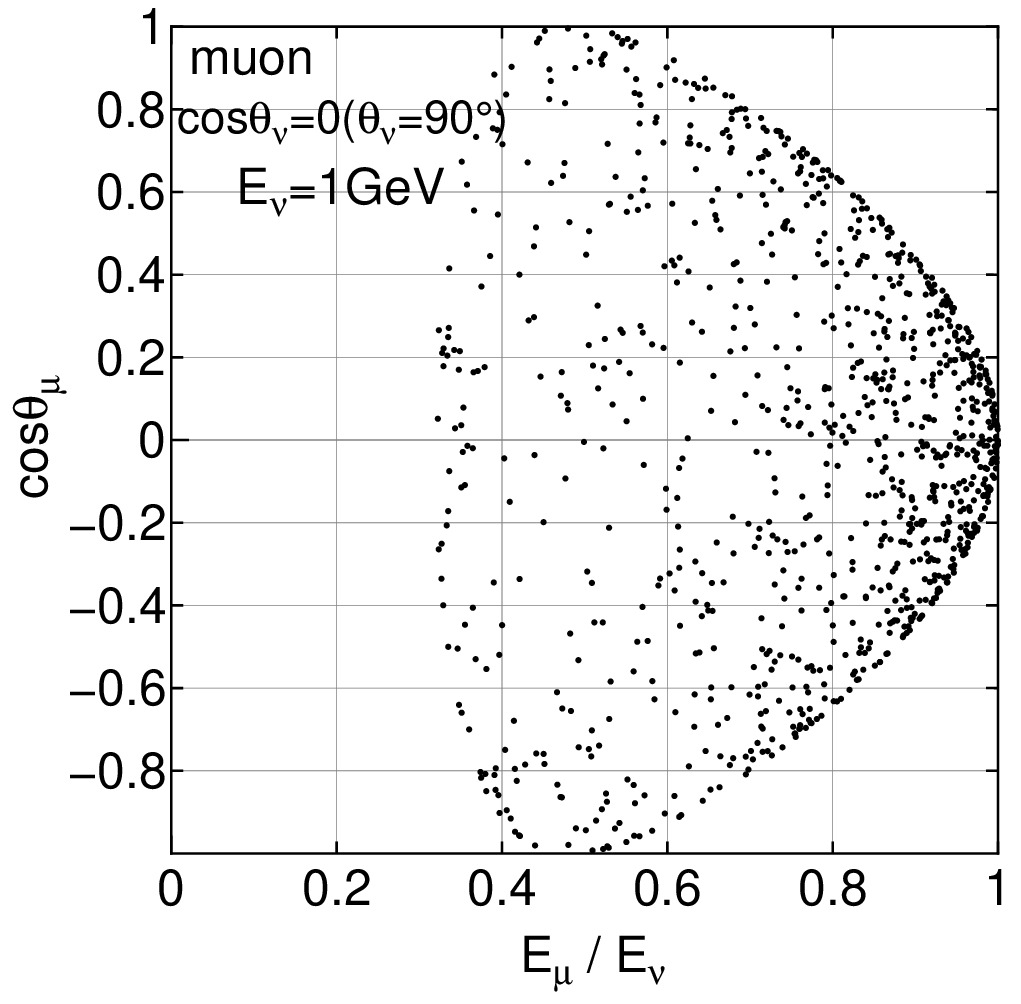}\hspace{1cm}
  \includegraphics{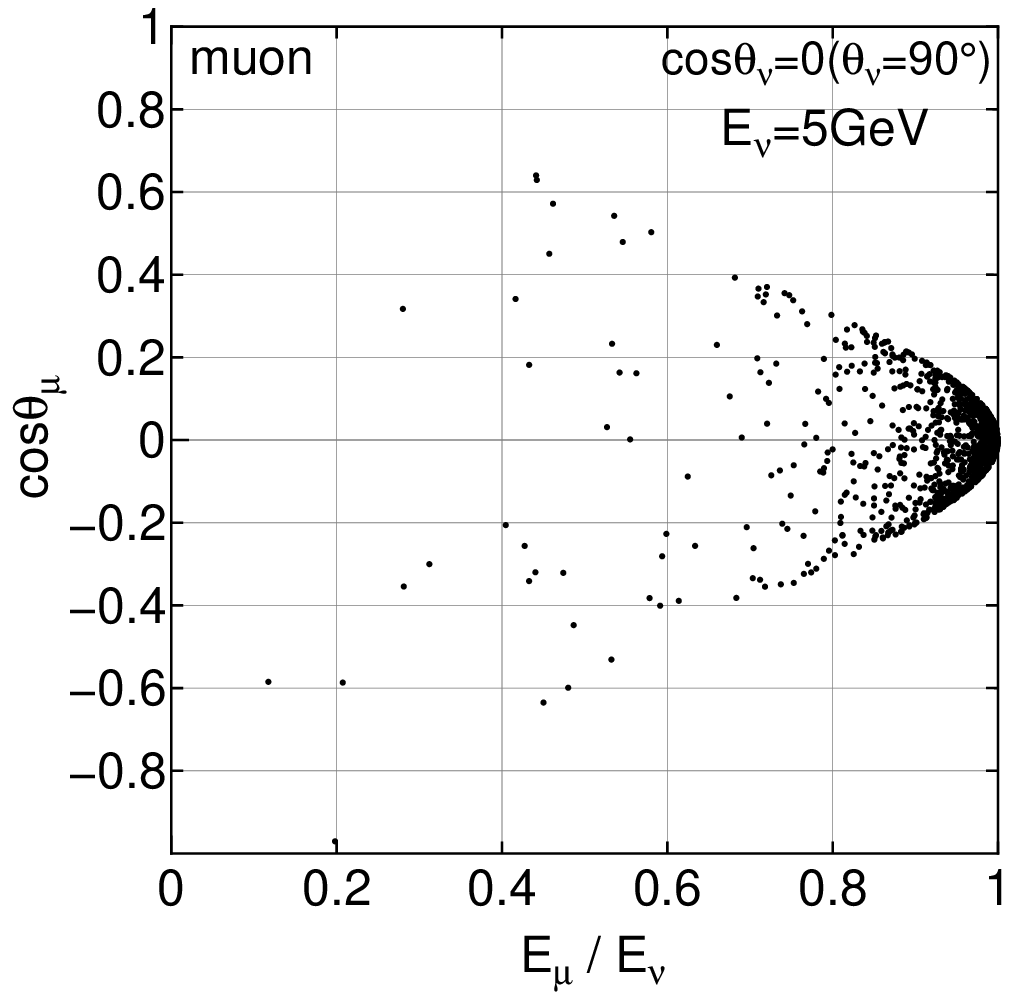}
}
\caption{
\label{figH005} 
The scatter plots of the fractional energies of the produced muons 
and their zenith angles for horizontally incident muon neutrinos 
with 0.5~GeV, 1~GeV and 5~GeV, respectively.
 The sampling number is 1000 in each case.
}
\end{center}
\vspace{0.5cm}
\hspace{2.5cm}(a)
\hspace{5.5cm}(b)
\hspace{5.5cm}(c)
\vspace{-0.3cm}
\begin{center}
\resizebox{\textwidth}{!}{%
  \includegraphics{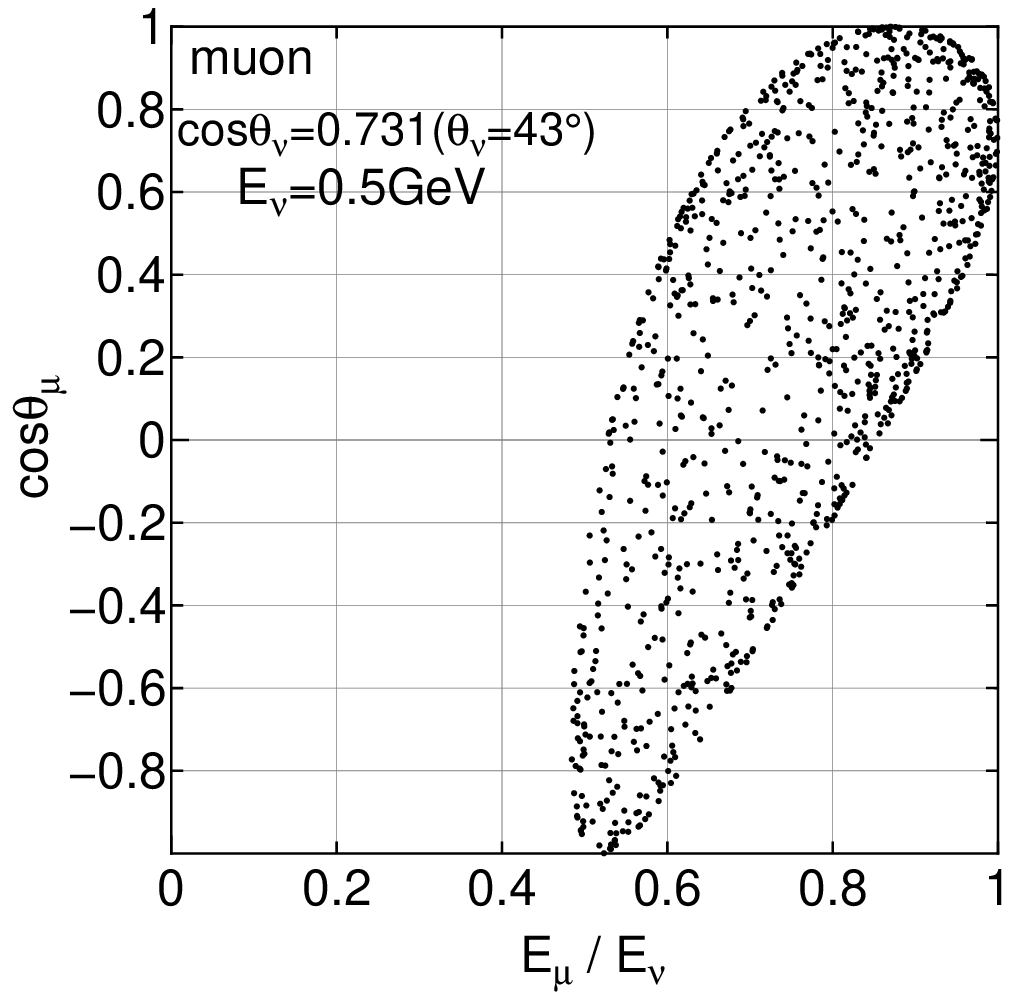}\hspace{1cm}
  \includegraphics{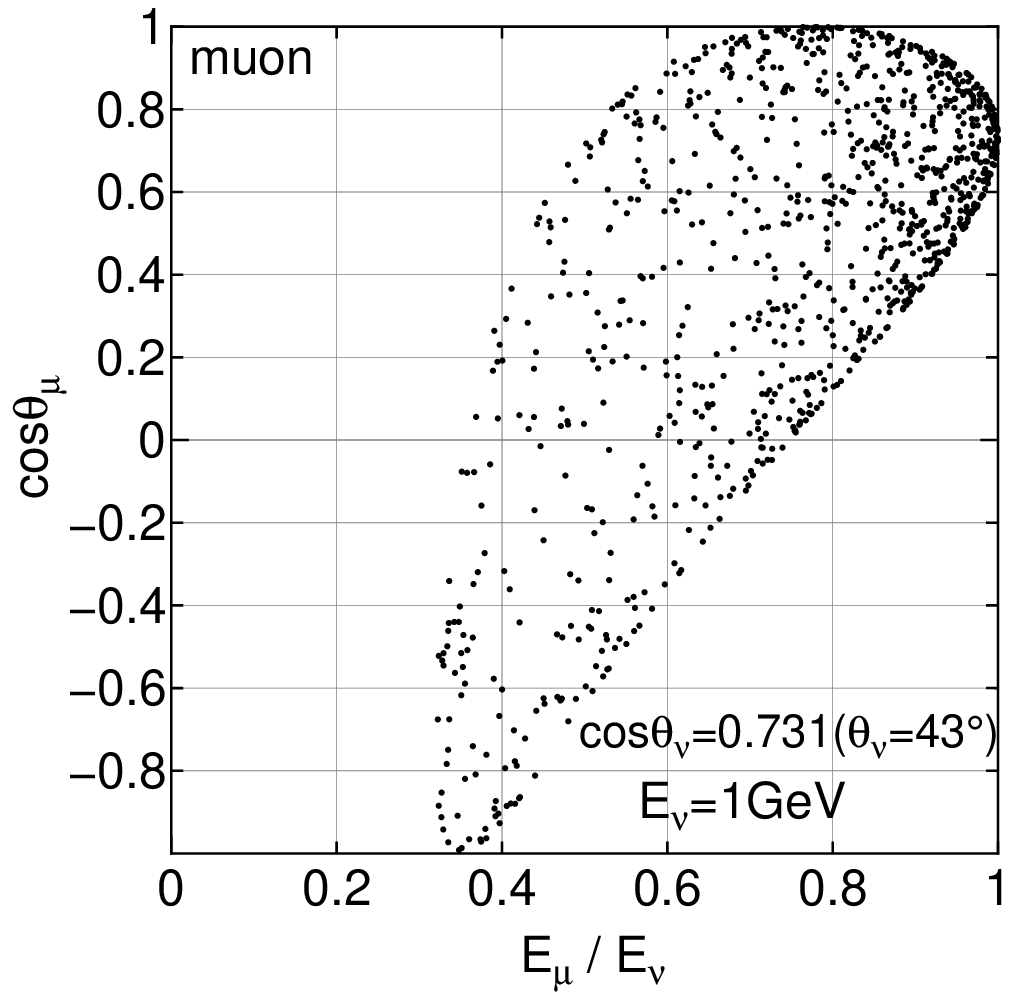}\hspace{1cm}
  \includegraphics{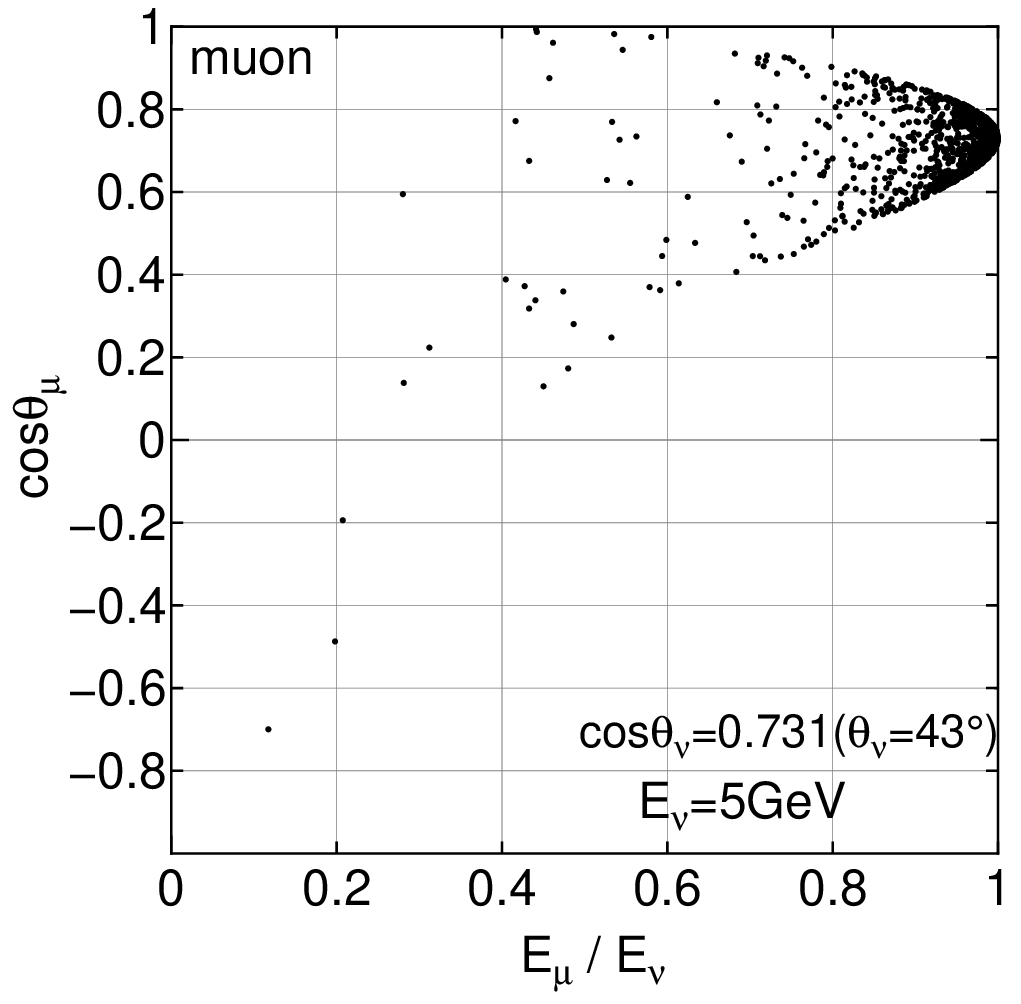}
  }
\caption{
\label{figH006} 
The scatter plots of the fractional energies of the produced muons 
and their zenith angles for diagonally incident muon neutrinos 
with 0.5~GeV, 1~GeV and 5~GeV, respectively.
 The sampling number is 1000 in each case.
}
\end{center}
\end{figure*} 

\subsection{
Dependence of the width of the zenith angle distribution of emitted 
leptons on their energy for different incident directions and energies 
of the neutrinos
}
  The detailed procedure for our Monte Carlo simulation is described in 
Appendix A. We give the scatter plots between the fractional energy of 
the emitted muons and their zenith angle for a definite zenith angle 
of the incident neutrino with different energies in 
Figures~\ref{figH004} to ~\ref{figH006}. 
In Figure~\ref{figH004}, we give the scatter plots for 
vertically incident neutrinos with different energies 0.5, 1 and 5~GeV. 
In this case, the relations between the emitted energies of the muons 
and their zenith angles are unique, which comes from the definition of 
the zenith angle of the emitted lepton\footnote{
The zenith angles of the particles concerned are measured from the 
downward direction.}.
However, the densities (frequencies of the event numbers) along each 
curve are different in position to position and depend on the energies of 
the incident neutrinos. Generally speaking, densities along the 
curves become greater toward $cos\theta_{\mu(\bar{\mu})}$= 1.
  In this case, $cos\theta_{\mu(\bar{\mu})}$ is never influenced by the 
azimuthal angle in the scattering by the definition$^2$.
On the contrary, it is shown in Figure~\ref{figH005} that the 
horizontally incident neutrinos give the widest zenith angle 
distribution for the emitted muons with the same fractional energies, 
an effect exclusively due to the azimuthal angle. The lower the energies 
of the incident neutrinos, the wider the spreads of the scattering 
angles of emitted muons $\theta_{\mu(\bar{\mu})}$, 
which leads to wider zenith angle distributions for the emitted muons and, 
therefore, we cannot estimate the directions of the incident neutrinos 
from the directions of the emitted leptons, especially in lower energies. 
As is easily understood from Figure~\ref{figH005}, the diagonally incident 
neutrinos give the intermediate zenith angle distributions for the 
emitted muons between those for vertically incident neutrinos and those 
for horizontally incident neutrinos.
It should be noted from the figures that the influence of the azimuthal 
angles in QEL on the cosines of the zenith angle of the incident neutrino 
exists for the most inclined neutrinos, even when the scattering angle 
due to QEL is small.

\begin{figure*}
\vspace{-1.0cm}
\hspace{2.5cm}(a)
\hspace{5.5cm}(b)
\hspace{5.5cm}(c)
\vspace{-0.3cm}
\begin{center}
\resizebox{\textwidth}{!}{%
  \includegraphics{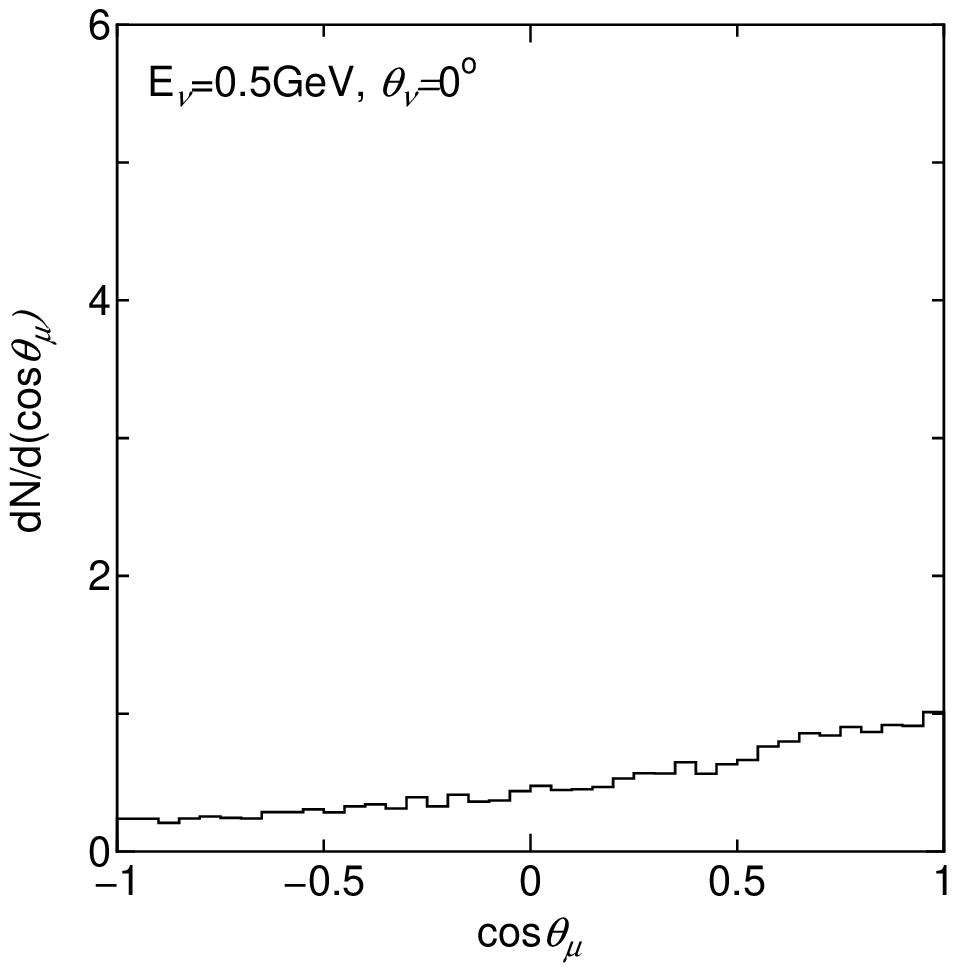}\hspace{1cm}
  \includegraphics{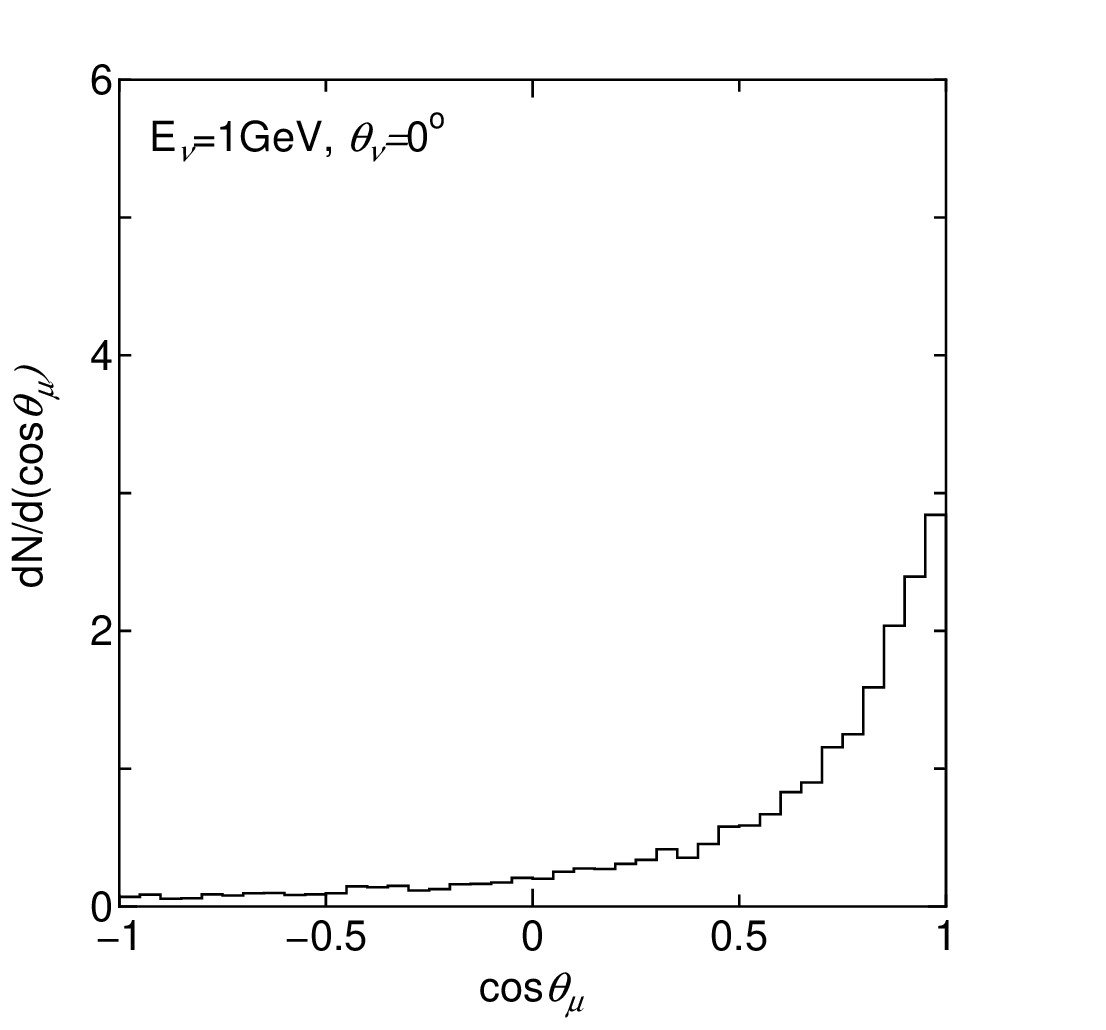}\hspace{1cm}
  \includegraphics{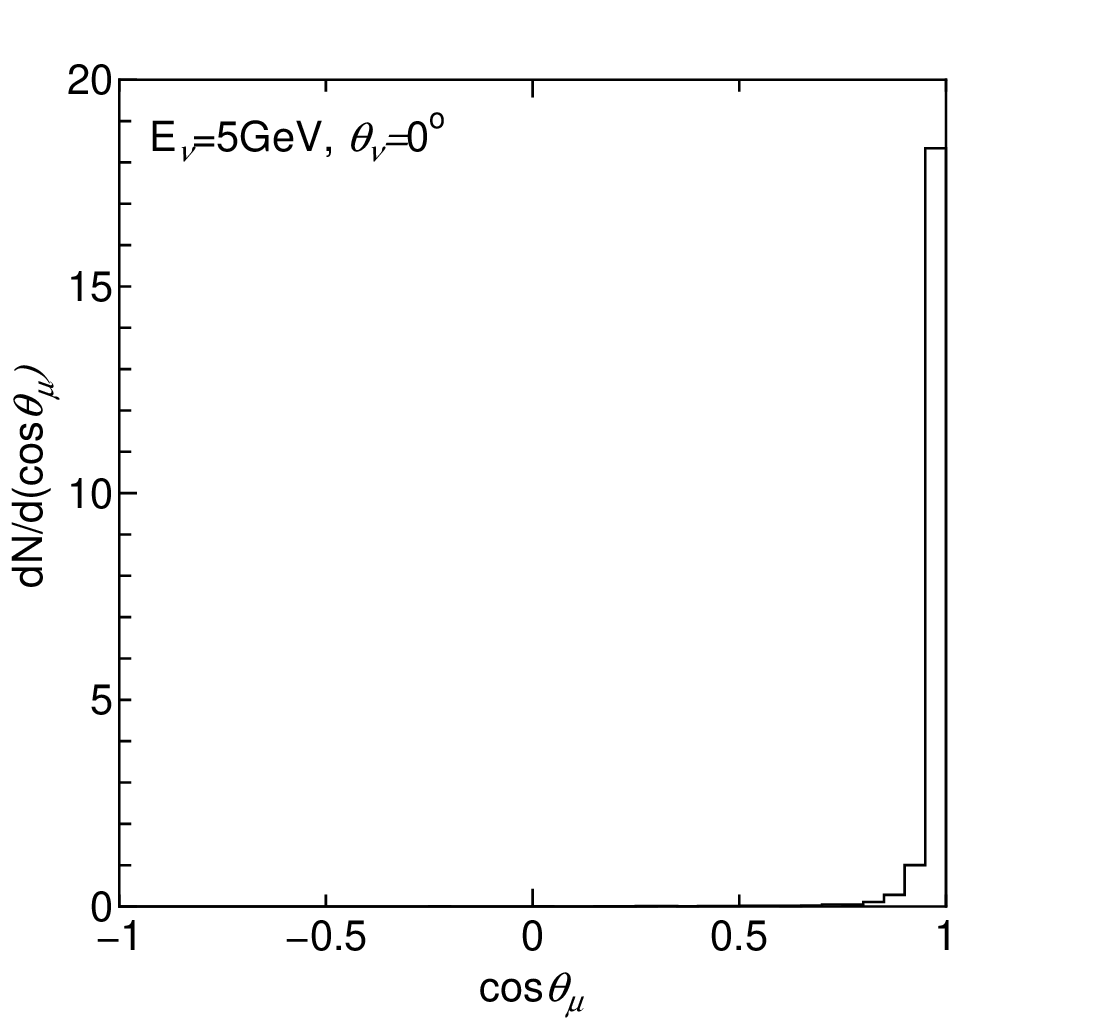}
  }
\caption{
The zenith angle distribution of the muons for vertically incident muon 
neutrinos with 0.5~GeV, 1~GeV and 5~GeV, respectively. The sampling 
number is 10000 in each case.
}
\label{figH007}
\end{center}
\vspace{0.5cm}

\hspace{2.5cm}(a)
\hspace{5.5cm}(b)
\hspace{5.5cm}(c)
\vspace{-0.3cm}
\begin{center}
\resizebox{\textwidth}{!}{%
 \includegraphics{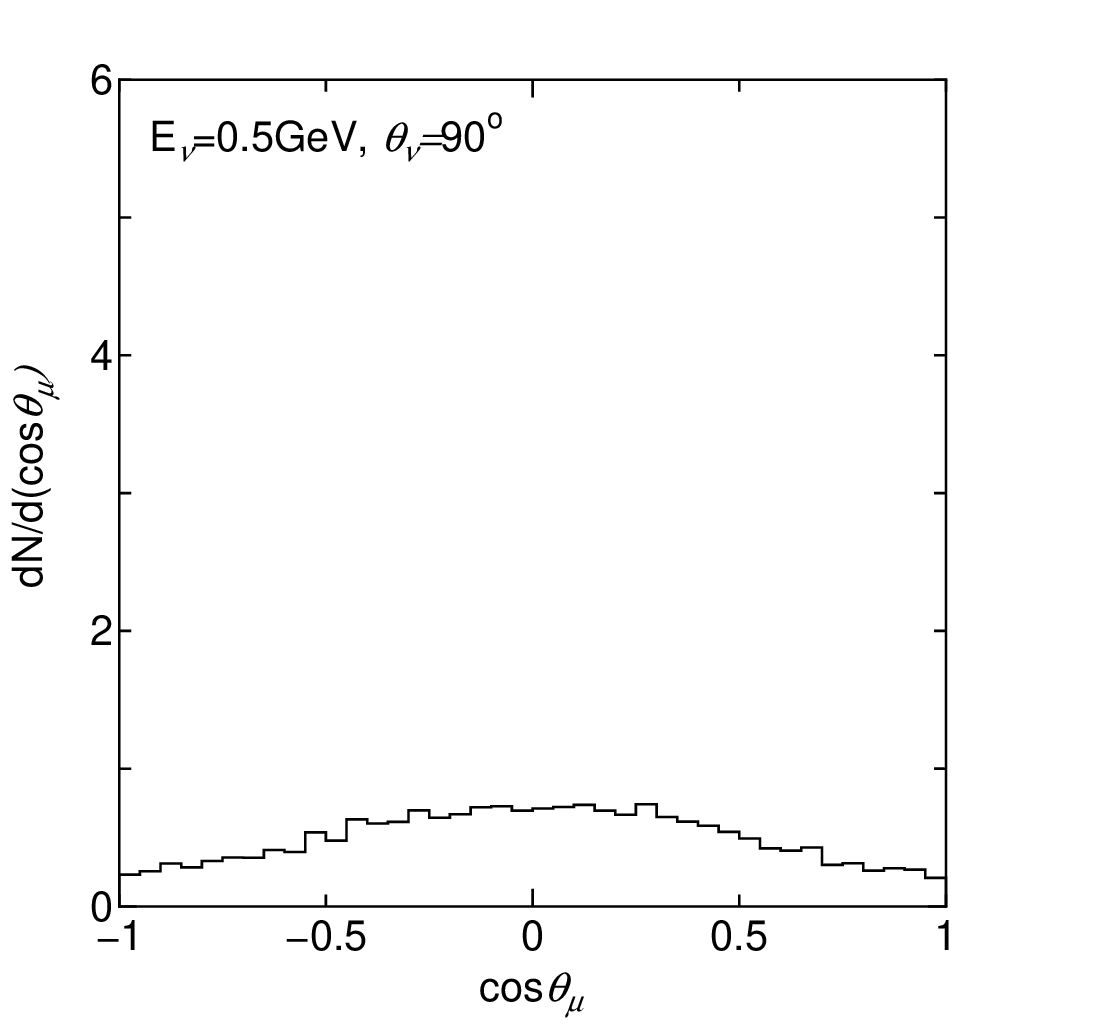}\hspace{1cm}
  \includegraphics{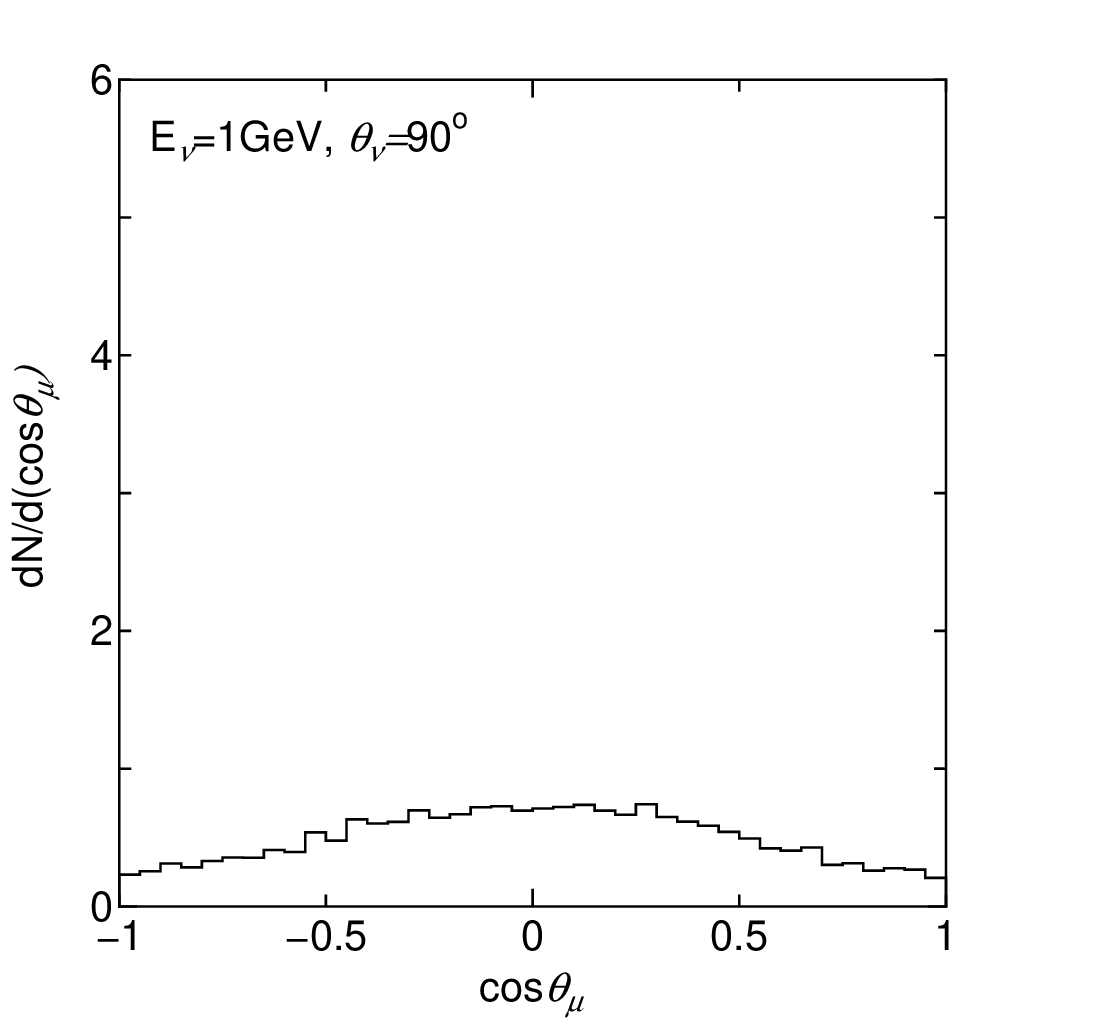}\hspace{1cm}
  \includegraphics{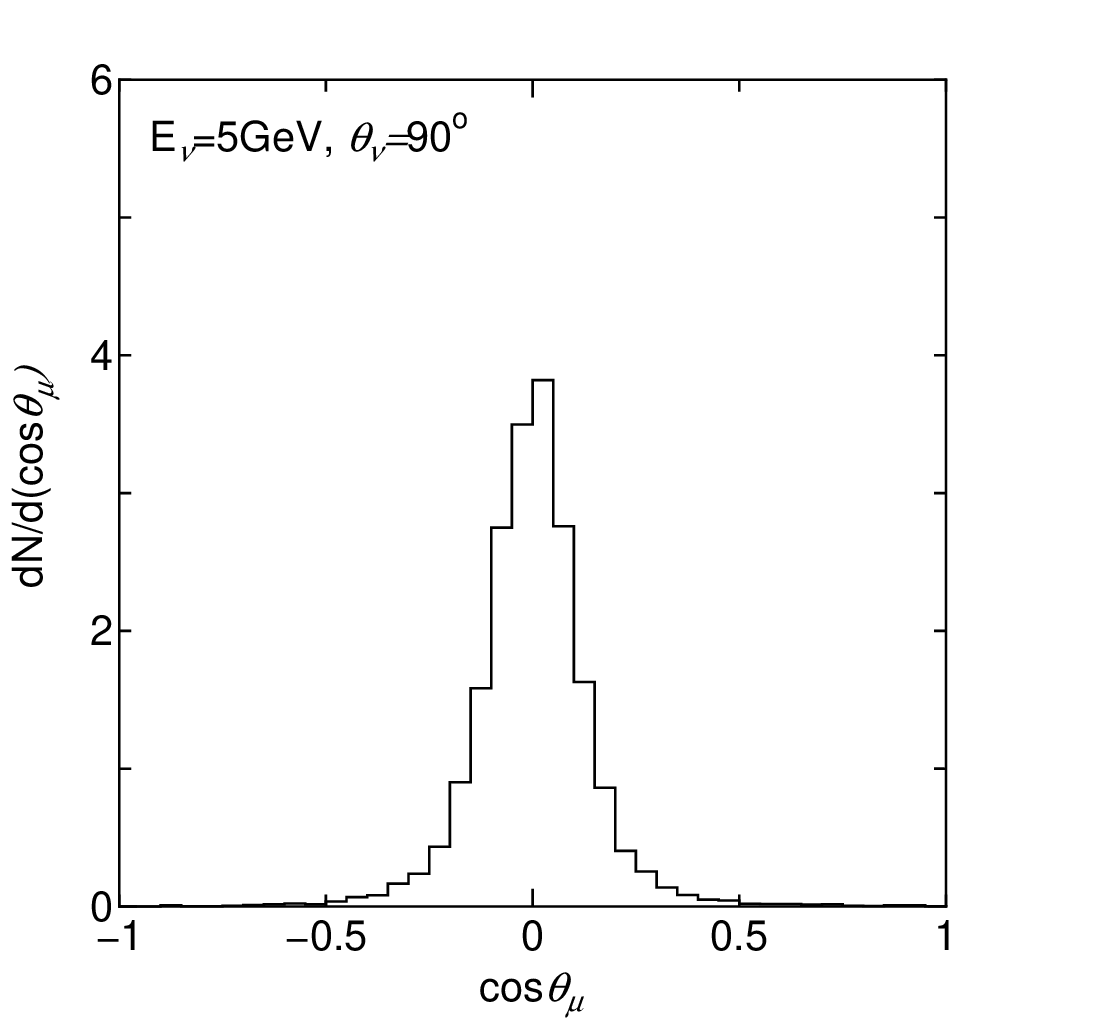}
  }
\caption{
The zenith angle distribution of the muons for horizontally incident muon 
neutrinos with 0.5~GeV, 1~GeV and 5~GeV, respectively. The sampling 
number is 10000 in each case.
}
\label{figH008}
\end{center}
\vspace{0.5cm}
\hspace{2.5cm}(a)
\hspace{5.5cm}(b)
\hspace{5.5cm}(c)
\vspace{-0.3cm}
\begin{center}
\resizebox{\textwidth}{!}{%
  \includegraphics{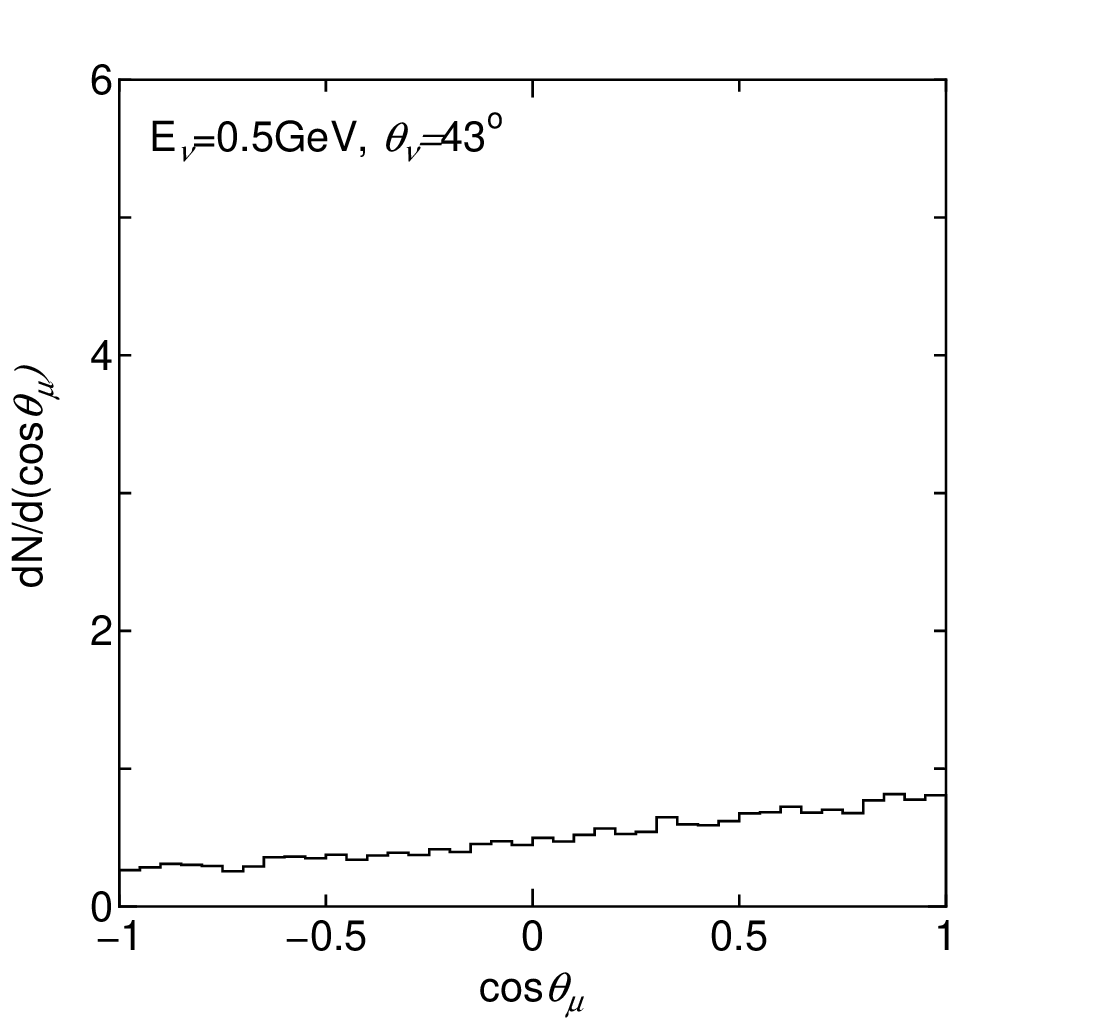}\hspace{1cm}
  \includegraphics{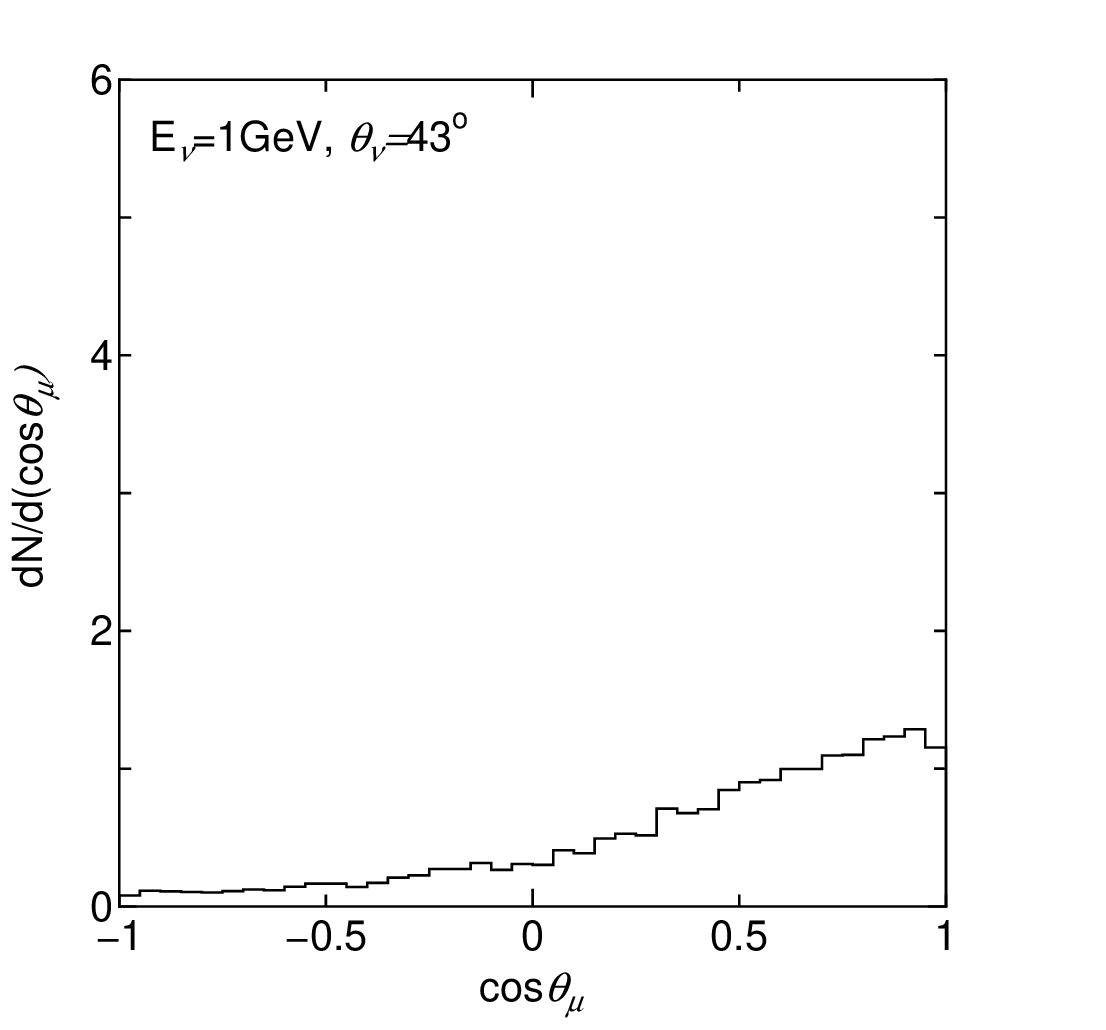}\hspace{1cm}
  \includegraphics{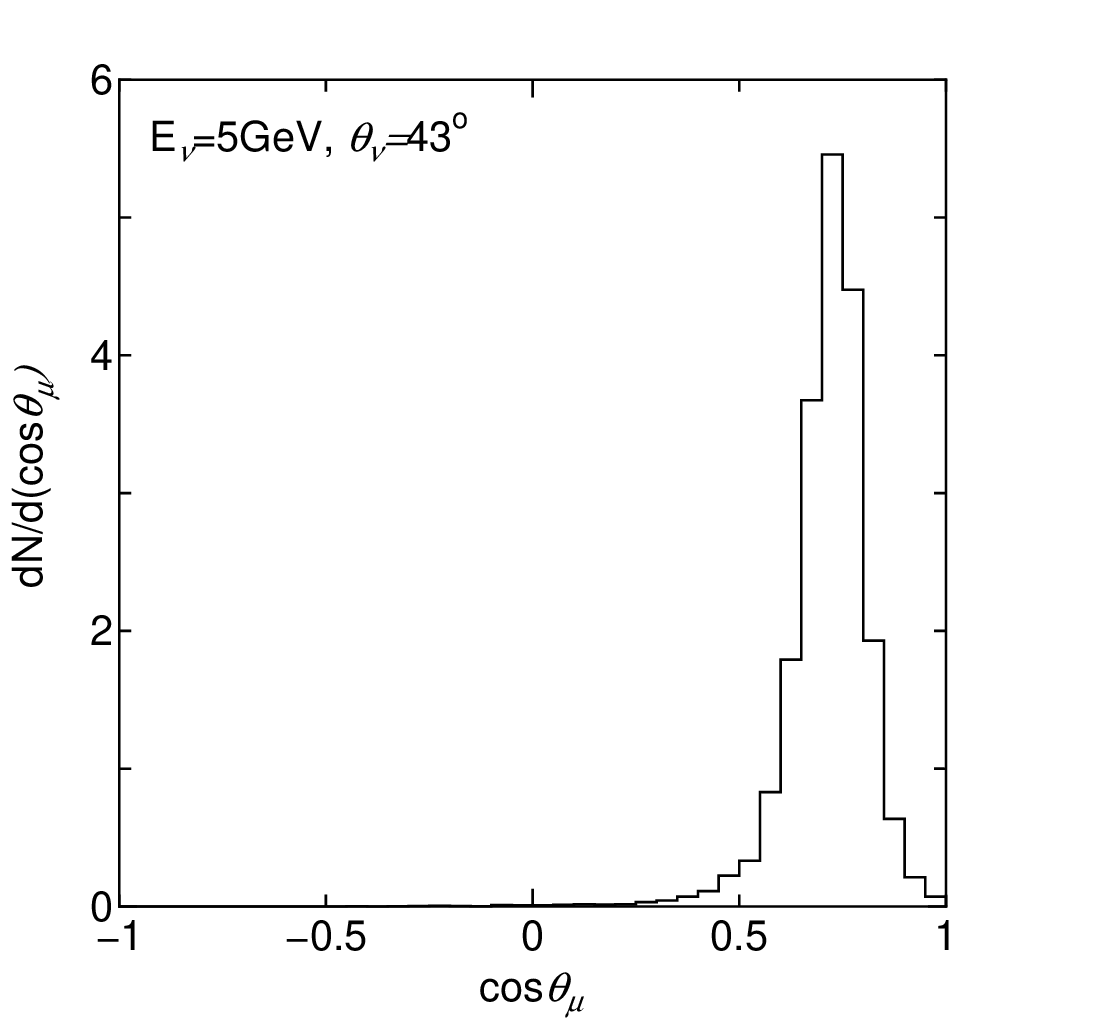}
  }
\caption{
The zenith angle distribution of the muons for diagonally incident muon 
neutrinos with 0.5~GeV, 1~GeV and 5~GeV, respectively. The sampling 
number is 10000 in each case.
}
\label{figH009}
\end{center}
\end{figure*} 

\subsection{
Zenith angle distribution of the emitted leptons for different incident directions and energies of 
the neutrinos
}

 In Figures~\ref{figH007} to \ref{figH009}, we show distributions of the 
zenith cosine of the emitted muons for different incident directions and 
different incident energies of neutrinos.
 In Figures~\ref{figH007}(a) to \ref{figH007}(c), we give the zenith 
cosine distributions of the emitted muons for the case of vertically 
incident neutrinos with different energies, $E_{\nu}$=~0.5, 1 and 5~GeV. 
Comparing the case for $E_{\nu}$=~0.5~GeV with that for $E_{\nu}$=~5~GeV, 
we see big difference between the two. The scattering angle of the 
emitted muon for 5 GeV neutrinos is relatively small (see also Table~1), 
so that the emitted muons keep roughly the same directions as their 
original neutrinos. In this case, the effect of their azimuthal angle on 
the resulting zenith angle is also smaller. However, in the case of 
$E_{\nu}$=~0.5~GeV, the dominant energy for the 
neutrino events which are fully contained in the detector, not a small 
percentage of the muons are emitted in the backward direction due to 
large angle scattering. The most frequent occurrence in the 
backward direction of the emitted muon appears in the horizontally 
incident neutrino as shown in Figures~\ref{figH008}(a) to 
~\ref{figH008}(c). In this case, the zenith angle distribution of the 
emitted muon should be symmetrical with regard to the horizontal 
direction. Comparing the case for 5 GeV with those both for 0.5~GeV and 
for 1~GeV, even 1~GeV incident neutrinos lose almost their memory of the 
incident direction. 
Figures~\ref{figH009}(a) to \ref{figH009}(c) for diagonally incident 
neutrinos tell us that the situation for the diagonal case lies between 
the case for the vertically incident neutrinos and that for horizontally 
incident ones. From Figures ~\ref{figH007}(a) to \ref{figH009}(c),
 it is clear that the scattering angles of emitted muons affect the 
determination of zenith angles of incident neutrinos. 
The effect is enhanced by their azimuthal angles, particularly for more 
inclined directions of the incident neutrinos.
\begin{figure*}
\vspace{-0.5cm}
\hspace{2cm}(a)
\hspace{5cm}(b)
\hspace{5cm}(c)
\vspace{-0.2cm}
\begin{center}
\resizebox{\textwidth}{!}{%
  \includegraphics{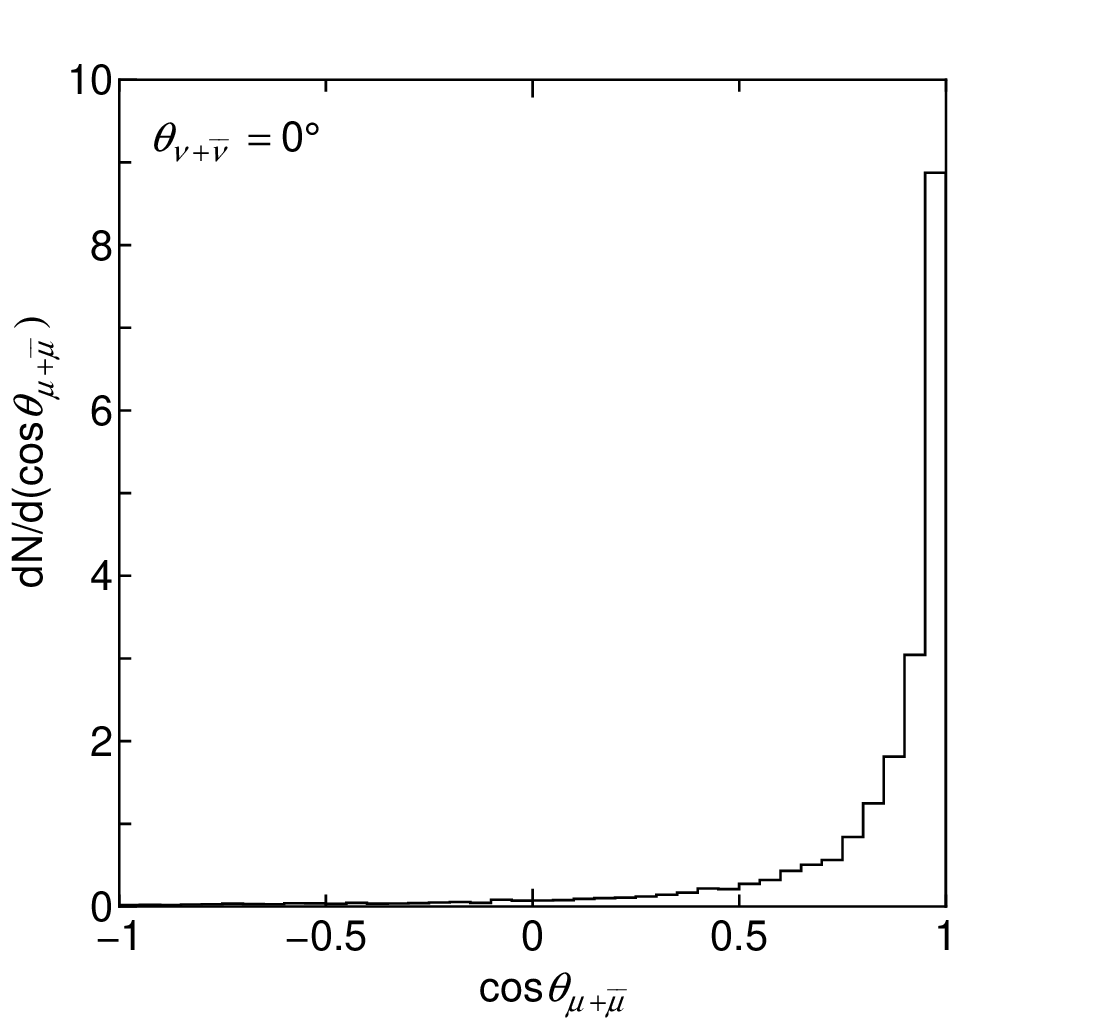}\hspace{1cm}
  \includegraphics{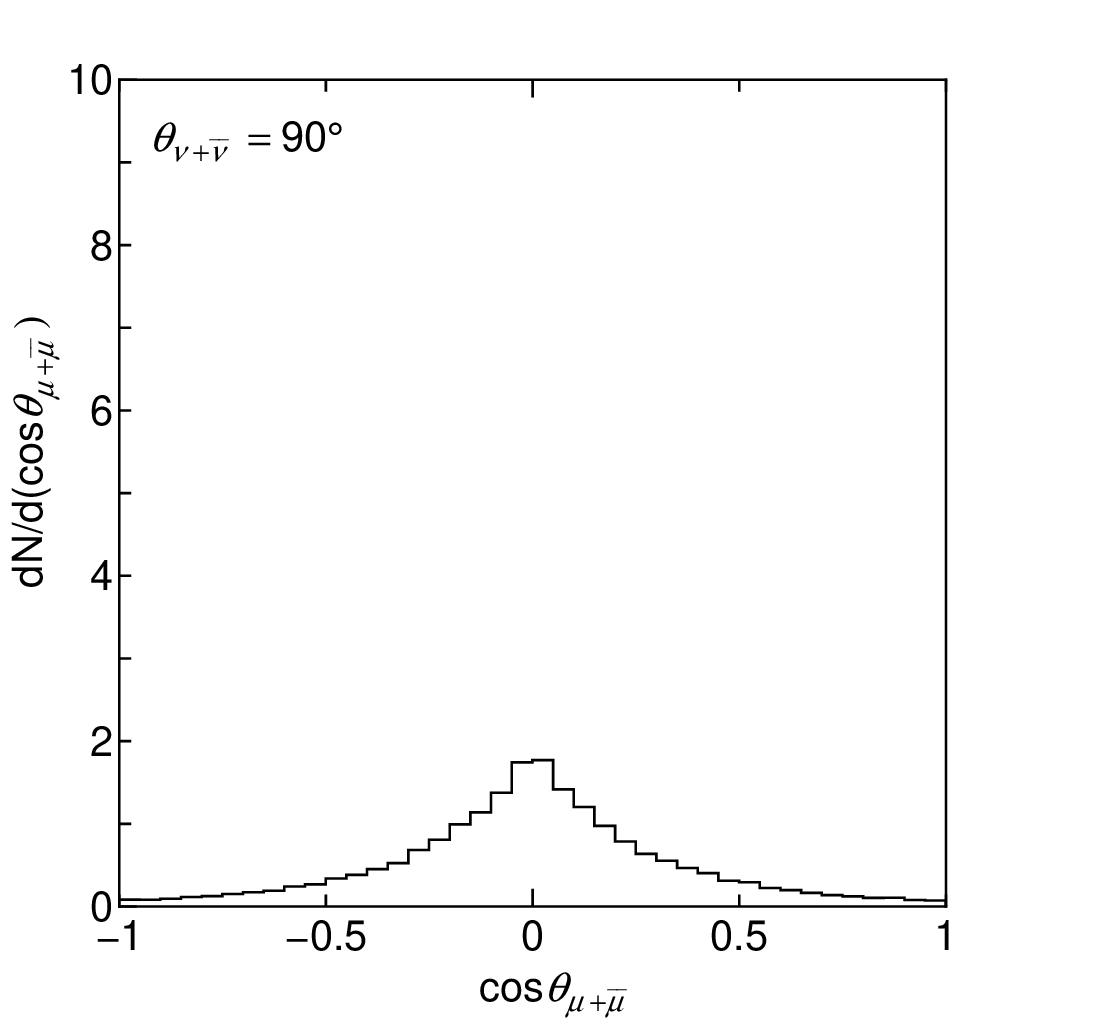}\hspace{1cm}
  \includegraphics{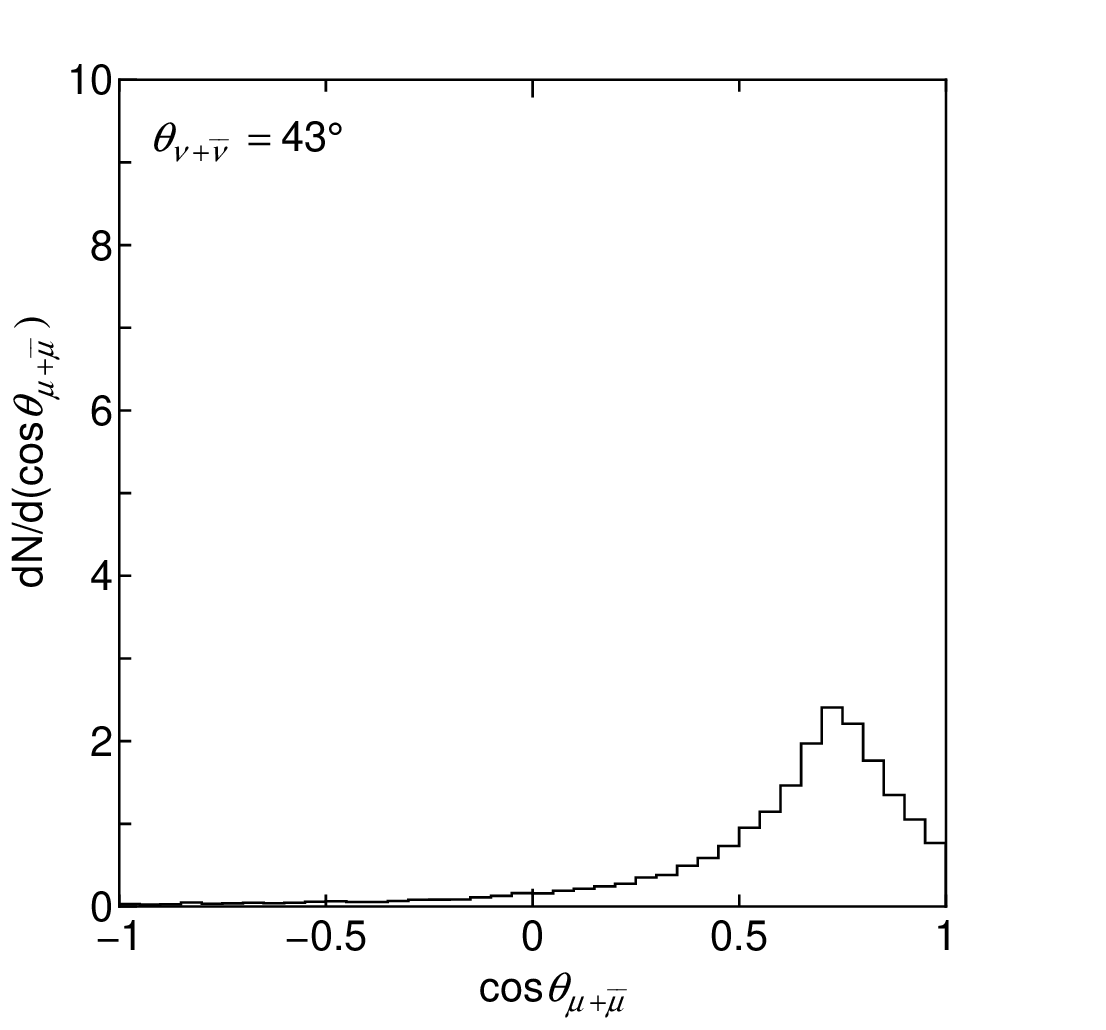}\hspace{1cm}
  }\\
 \end{center}
\caption{ 
The zenith angle distribution for the sum of $\mu^-$ and $\mu^+$ for
incident neutrinos, $\bar{\nu}$ and $\nu$ for the vertical, horizontal 
and diagonal directions, respectively (see Figure~\ref{figH003}). 
The overall neutrino spectra at the Kamioka site are taken into account. The sampling number is 10000 in each case. 
}
\label{fig:H010}
\end{figure*} 

\subsection
{Zenith angle distribution of Fully Contained Events for a given zenith 
angle of the incident neutrino, taking their energy spectrum into account
}

 In the previous subsections we discuss the relation between the zenith 
angle distribution of the incident neutrino and that of the emitted muons 
produced by the neutrinos for three different directions and three 
different energies of incident neutrinos. In order to apply our 
inspection to the real experiment using the underground detectors,
 we must consider the effect of the energy spectrum of the incident 
neutrinos.  
Here, we adopt Honda's spectrum\cite{Honda} for incident neutrino 
energy. Details of the Monte Carlo simulations for this purpose are given 
in Appendix B.
In Figure~\ref{fig:H010}, we give the zenith angle distributions of the 
sum of $\mu^+$($\bar{\mu}$) and $\mu^-$  and  for a given incident 
zenith angle of $\nu_{\bar{\mu}}$ and $\nu_{\mu}$, taking into account 
the different primary neutrino energy spectra for 
different directions at the underground detector at the Kamioka site
\cite{Honda}.  It is clear from these figures that 
the zenith angles of the {\it Fully Contained Events} are greatly 
influenced by the scattering angle of emitted leptons.
 The effect becomes much more in case of inclined events if we take the 
azimuthal angle into consideration.

\newpage
\section
{The correlation between $cos\theta_{\nu(\bar{\nu})}$ (or 
$L_{\nu(\bar{\nu})}$) and
$cos\theta_{\mu(\bar{\mu})}$ (or $L_{\mu(\bar{\mu})}$), and the 
correlation between $E_{\nu(\bar{\nu})}$  and $E_{\mu(\bar{\mu})}$
}
In this section and section~5, we give a part of the results obtained 
from the $L/E$ analysis for QEL events among the 
{\it Fully Contained Events} with our Computer Numerical Experiment.
 For the execution of our experiment, we borrow necessary data from the 
Super-Kamiokande Collaboration\cite{Ashie2}, the detector geometry, 
the duration of 1489.2 live days, oscillation parameters,
$sin^2 2\theta = 1.0$ and $\Delta m^2 = 2.4\times 10^{-3}\rm{eV^2}$,
and the incident neutrino spectrum\cite{Honda}.
Here, we examine $cos\theta_{\nu(\bar{\nu})}$ (or $L_{\nu(\bar{\nu})}$) 
and $cos\theta_{\mu(\bar{\mu})}$ (or $L_{\mu(\bar{\mu})}$), 
and the correlation between $E_{\nu(\bar{\nu})}$ and $E_{\mu(\bar{\mu})}$ 
in the QEL events among the {\it Fully Contained events}, 
where $cos\theta_{\nu(\bar{\nu})}$, $cos\theta_{\mu(\bar{\mu})}$,
$L_{\nu(\bar{\nu})}$ and  $L_{\mu(\bar{\mu})}$
are the zenith cosine of the incident neutrino and that of the emitted 
muon, the corresponding flight path length of the incident neutrino and 
that of the emitted muon, respectively.
\\

In the following and subsequent sections, we examine the validity of 
Eq.(2) in the QEL events among the {\it Fully Contained Events}.
There are two possibilities in order that Eq.(2) is valid.

\begin{itemize}
\item [CASE A]: The relations of both $L_{\nu} \approx L_{\mu}$and
$E_{\nu} \approx E_{\mu}$  hold good.
\item [CASE B]: In spite of the failure of Case A, Eq.(2) holds due to unknown reasons.
\end{itemize}
First of all, we examine CASE A.

\subsection
{The Correlation between $cos\theta_{\nu(\bar{\nu})}$ and 
$cos\theta_{\mu(\bar{\mu})}$
}
Here, we examine the correlation between $cos\theta_{\nu}$
 and $cos\theta_{\mu}$ for the neutrino events concerned (QEL events) by 
the computer numerical experiment assuming live days same to the real 
ones of the underground experiment\cite{Ashie2} and the incident neutrino 
energy spectrum as a function of zenith angle as in \cite{Honda}. The 
details are given in Appendix C.
 In order to obtain the zenith angle distribution of the emitted leptons 
arising from the incident neutrinos, we divide the range of cosine of the 
zenith angle of the incident neutrino into twenty regular intervals from 
 $cos\theta_{\nu}=-1$ (downward) to $cos\theta_{\nu}=1$ (upward).
 For given interval of $cos\theta_{\nu}$, we carry out the exact 
Monte Carlo simulation, starting to sample neutrino events in a 
stochastic manner from the incident neutrino energy spectrum for given   
$cos\theta_{\nu}$, the cosine of the zenith angles of the incident 
neutrinos, and obtain finally $cos\theta_{\mu}$, the cosine of the zenith 
angles of the emitted leptons.

\begin{figure*}
\begin{center}
\vspace{-1.5cm}
\rotatebox{90}{%
\resizebox{0.5\textwidth}{!}{%
  \includegraphics{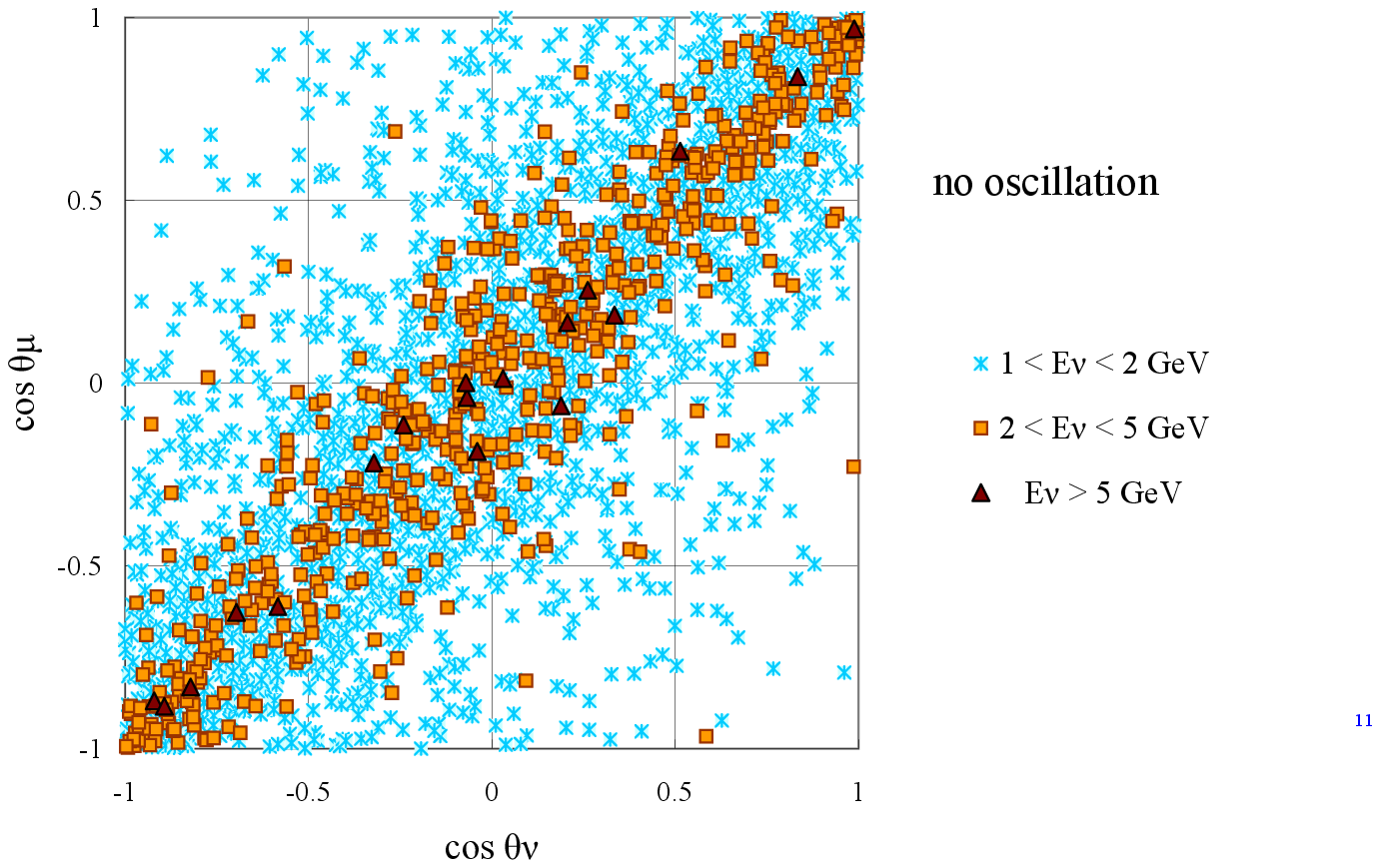}
  }}
\vspace{-1.5cm}
\caption{The correlation diagram between $\cos{\theta}_{\nu}$ and $\cos{\theta}_{\mu}$ for null oscillations
for different neutrino energy regions.
}
\label{figR011} 
\vspace{-0.2cm}
\rotatebox{90}{%
\resizebox{0.5\textwidth}{!}{%
  \includegraphics{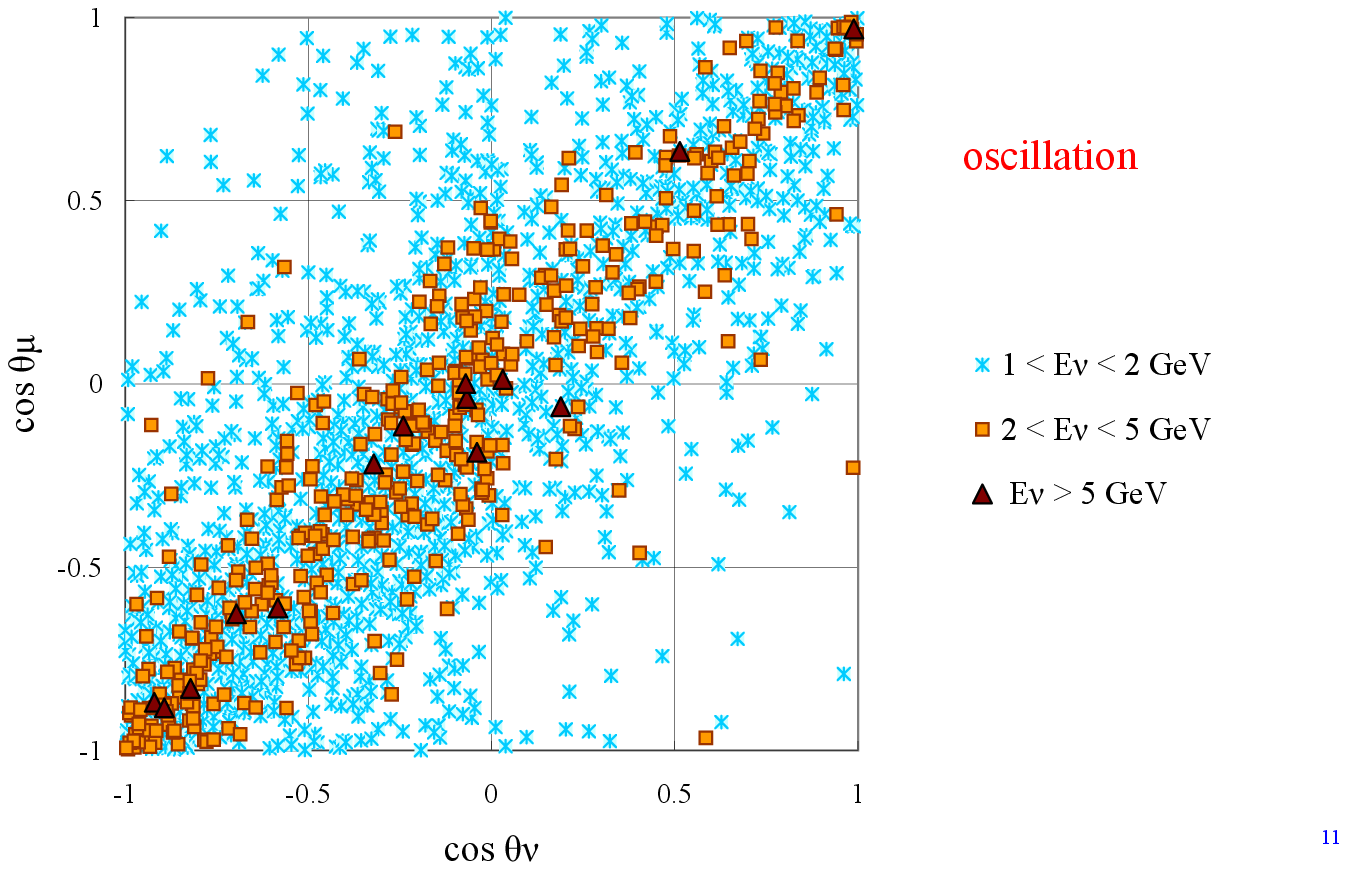}
  }}
\vspace{-1.5cm}
\caption{The correlation diagram between $\cos{\theta}_{\nu}$ and $\cos{\theta}_{\mu}$ for oscillations
for different neutrino energy regions.}
\label{figR012}
\end{center}
\end{figure*} 
\begin{figure*}
\begin{center}
\vspace{-1.5cm}
\rotatebox{90}{%
\resizebox{0.5\textwidth}{!}{%
  \includegraphics{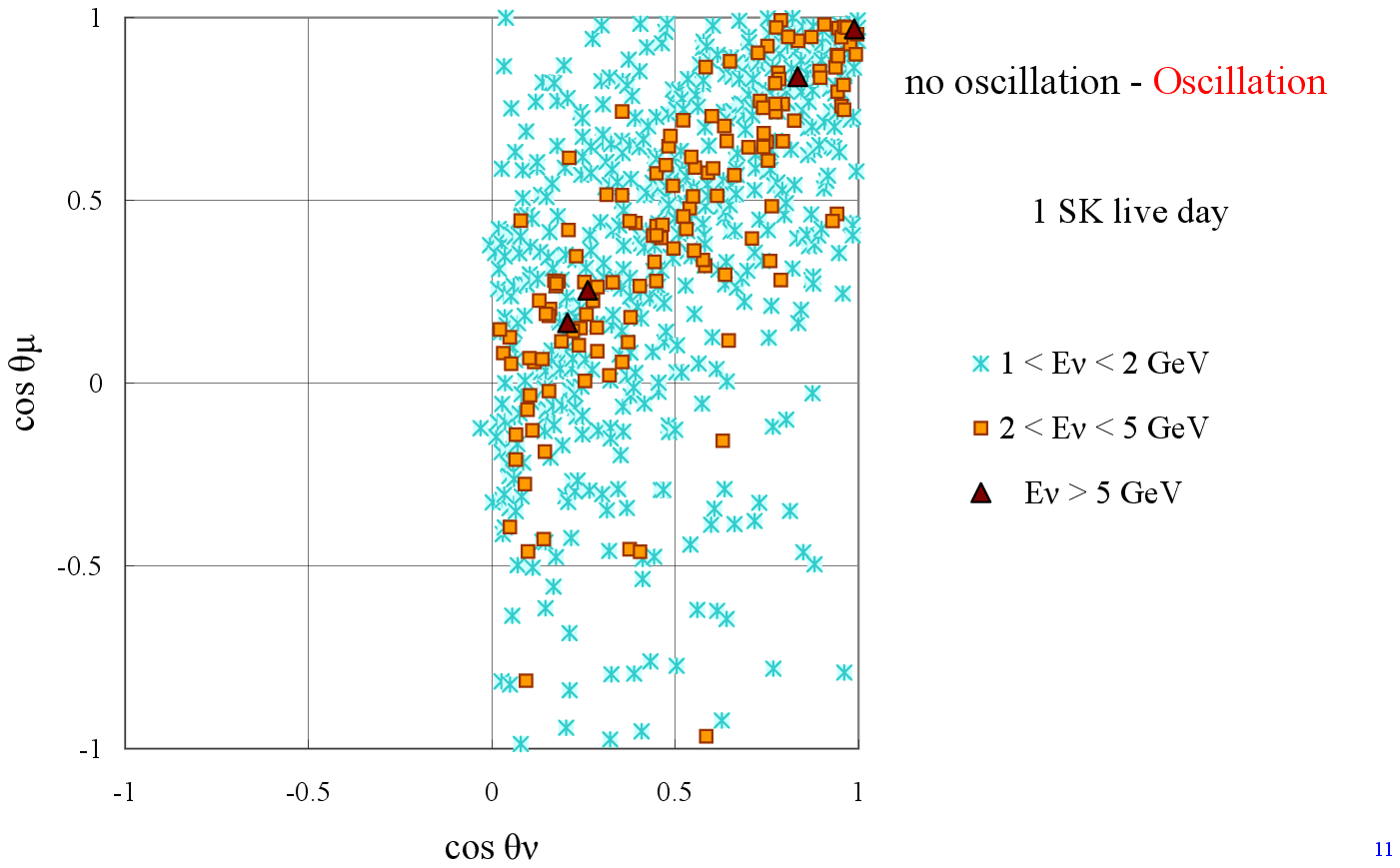}
  }}
\vspace{-1.5cm}
\caption{The correlation diagram between $\cos{\theta}_{\nu}$ and $\cos{\theta}_{\mu}$ for those events which exist for null oscillations
but disappear due to oscillations for different neutrino energy regions.}
\label{figR013} 

\end{center}
\end{figure*} 

\subsubsection
{The case without neutrino oscillations}  
In Figure~\ref{figR011} we give the correlation diagram between
$\cos{\theta}_{\nu}$ and $\cos{\theta}_{\mu}$ in case of no neutrino 
oscillation for 1489.2 live days. 

In the figure, we plot data using different symbols for three different 
primary energies,  $E_{\nu}>$~5~GeV , 2~$<E_{\nu}<$~5~GeV  and 
 1~$<E_{\nu}<$~2~GeV, in order to examine the influence of the incident 
energies on the correlation between $\cos{\theta}_{\nu}$ and 
$\cos{\theta}_{\mu}$. 
The neutrino events of $E_{\nu}>$~5~GeV are distributed roughly on the 
line $\cos{\theta}_{\nu} = \cos{\theta}_{\mu}$.
 This reflects the fact that the emitted leptons in such higher energy 
interactions are produced with smaller scattering angles 
so that they are less influenced by their azimuthal angles 
(see Table~1, Figure~\ref{figH002}, Figures~\ref{figH005}(c) to
\ref{figH006}(c)).
 In this energy region the directions of the incident neutrinos can be 
approximately determined by those of the emitted muons. 
The neutrino events of 2~$<E_{\nu}<$~5~GeV are also distributed along 
the line $\cos{\theta}_{\nu} = \cos{\theta}_{\mu}$, 
but the correlation is weaker than in the case of neutrino events of 
$E_{\nu}>$~5~GeV. 
In this energy region the direction of the incident neutrinos determined 
by those of the emitted muons has larger uncertainties. The 
influence of such larger uncertainty on the $L/E$ distribution must be 
carefully examined. In contrast to these 
higher energy ranges, $E_{\nu}>$~2~GeV, the neutrino events of lower 
energies, 1~$<E_{\nu}<$~2~GeV, are scattered out widely around the 
line $\cos{\theta}_{\nu} = \cos{\theta}_{\mu}$. 
The fact that the relation 
$\cos{\theta}_{\nu} \approx \cos{\theta}_{\mu}$ does not hold, 
as seen in the figure, plays a decisively important role in the $L/E$ 
analysis (see section~4.2). 
The influence of the 
emitted muons' directions on determining those of the incident neutrinos 
will be carefully examined in the subsequent papers, because it is 
closely related to the maximum oscillation in the survival probability.

It is interesting to classify the events in Figure~\ref{figR011} into the 
following four sectors with the regard to the origin
  ($\cos{\theta}_{\nu}=0,  \cos{\theta}_{\mu}=0$).
  
\begin{itemize}  
\item [(A)] The first sector where 
$\cos{\theta}_{\nu}>0$ and $\cos{\theta}_{\mu}>0$. In this sector we 
recognize that the incident 
neutrinos go upward ($\cos{\theta}_{\nu}>0$) and the emitted leptons 
also go upward ($\cos{\theta}_{\mu}>0$) in the interaction.  
Namely, the scattered leptons are produced in the forward direction by the upward neutrinos (see footnote 2).
\item[(B)] The second sector where $\cos{\theta}_{\nu}<0$  
and $\cos{\theta}_{\mu}>0$. In this sector we recognize that the incident 
neutrinos go downward ($\cos{\theta}_{\nu}<0$) but the emitted leptons go 
upward ($\cos{\theta}_{\mu}>0$) in the interaction. 
Such situations may occur due to two different causes. One is that the 
emitted leptons are scattered in the backward directions (backward 
scattering, see Figures~\ref{figH001} and \ref{figH002}), 
while the other is that the emitted leptons are scattered forward by the 
downward neutrinos, but they look like backward scattering due to the 
azimuthal angle effect (see Figure~\ref{figH003}(b) 
and \ref{figH003}(c)). 
\item[(C)] The third sector where $\cos{\theta}_{\nu}<0$ and
$\cos{\theta}_{\mu}<0$. In this sector we recognize that the incident 
neutrinos go downward and the emitted leptons are scattered forward in 
the interaction.
\item[(D)] The fourth sector where  $\cos{\theta}_{\nu}>0$ and 
$\cos{\theta}_{\mu}<0$. In this sector we recognize that the incident 
neutrinos go upward ($\cos{\theta}_{\nu}>0$) and the emitted leptons go 
downward ($\cos{\theta}_{\mu}<0$) in the interaction. 
These situations also may occur due to different causes. One is that the 
emitted leptons are scattered backward (backward scattering, see Figures ~
\ref{figH001} and \ref{figH002}) and the other is that the emitted 
leptons are scattered forward, but they looks like backward scattering 
due to the azimuthal angle effect (see Figure~\ref{figH003}(b) 
and \ref{figH003}(c)). 
\end{itemize}

We can see in Figure~\ref{figR011} that the distribution of the neutrino 
events in the first sector and that in the third sector is essentially 
symmetrical and that in the second sector and that in the fourth sector is also symmetrical. 
Such symmetry is easily understood from the fact that the mean free paths 
of the neutrino interactions in the energy region smaller than 10~GeV, 
which corresponds to the maximum energy for 
{\it Fully Contained Events}, is far larger than the diameter of the 
Earth and consequently neutrino fluxes for the interaction are 
independent of the thickness traversed through the Earth by the incident 
neutrinos. The reason for the smaller number of the events in the second 
sector (or the fourth sector), compared with that in the third sector 
(or the first sector) is that the probability of backward scattering of 
muons is smaller than that of forward scattering. 
The existence of such symmetries gives certain evidence that our Computer 
Experiment is carried out in a correct manner.
 
\subsubsection
{ The case with neutrino oscillations}  
 In Figure~\ref{figR012}, we give the corresponding correlation diagram 
for the case with neutrino oscillations, which should be compared with 
Figure~\ref{figR012}(no oscillations).  
In the case of the presence of neutrino oscillations, the symmetries 
between the first and third sectors, and those between the second and 
fourth sectors, are lost due to the neutrino oscillation effect. 
This is because the neutrino flux of the upward 
neutrinos is reduced compared with that of the downward neutrinos due 
to the survival probability (See Eq.(\ref{eqn:1})). 
 Of course, the destruction of these symmetries depends on the values of 
the oscillation parameters, $\Delta m^2$ and $sin^2 2\theta$
\footnote
{In our computer numerical experiments, we adopt the rejection method in 
Monte Carlo manner to obtain neutrino events with oscillation.}. 
In Figure~\ref{figR013}, we show the disappeared events which are the 
result of neutrino oscillations, i.e. the neutrino events resulting 
from subtraction of Figure~\ref{figR012} from Figure~\ref{figR011}). 
It should be noticed from the figure that the disappeared events are 
concentrated in the first sector and the fourth sector and that almost no 
disappeared event are to be found in the second sector and the third 
sector. 
The reasons are as follows: The neutrino events in the second sector and 
the third sector are due to the downward neutrinos which suffer almost no 
oscillation effect. Consequently, there are almost no disappeared events, 
as shown in Figure~\ref{figR012}. On the contrary, neutrino events in the 
first sector and the fourth sector are due to the upward neutrinos, which 
can be affected by the neutrino oscillation.  
The neutrino events in the first sector are due to the forward scattering 
and those in the fourth sector are due to the backward scattering and 
consequently the number of the neutrino event in the first sector is 
larger than that in the fourth sector, because the probability of forward 
scattering is larger than for backscattering (see Figure~\ref{figH002}).

This situation is closely related to the adoption of the specified 
neutrino oscillation parameters of $\Delta m^2$ and $sin^2 2\theta$
\footnote
{In our numerical experiment, we adopt
$\Delta m^2 = 2.4\times 10^{-3}\rm{ eV^2}$ and 
$sin^2 2\theta=1.0$\cite{Ashie2}.}.
The influence of the energies of the incident neutrinos on the 
correlation between 
$cos\theta_{\nu(\bar{\nu})}$ and 
$cos\theta_{\mu(\bar{\mu})}$
with oscillations is essentially the same as in the case without 
oscillations. The difference comes from the event rate due to neutrino 
oscillation effect.

\begin{figure}
\begin{center}
\vspace{-2cm}
\hspace*{-2.5cm}
\rotatebox{90}{%
\resizebox{0.5\textwidth}{!}{%
  \includegraphics{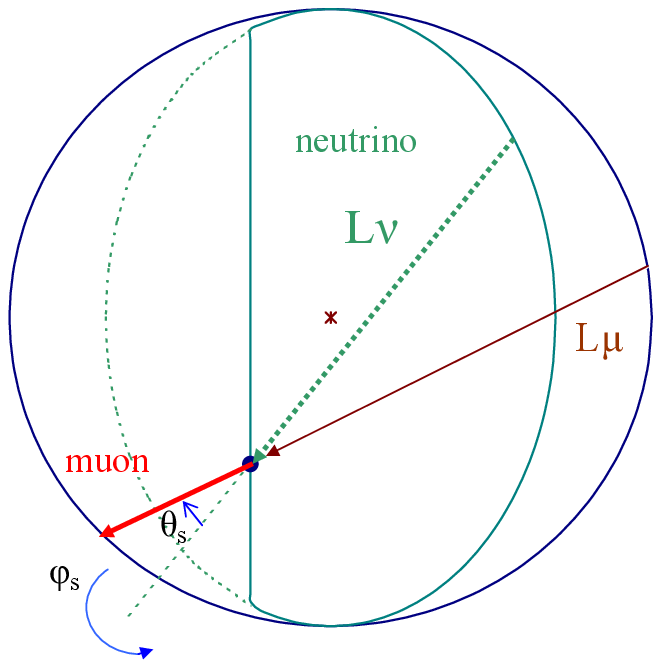}
}}
\vspace{-2cm}
\caption{Schematic view of relations among
$L_{\nu}$, $L_{\mu}$, $\theta_s$ and $\phi_s$
.}
\label{figR014}
\resizebox{0.5\textwidth}{!}{%
  \includegraphics{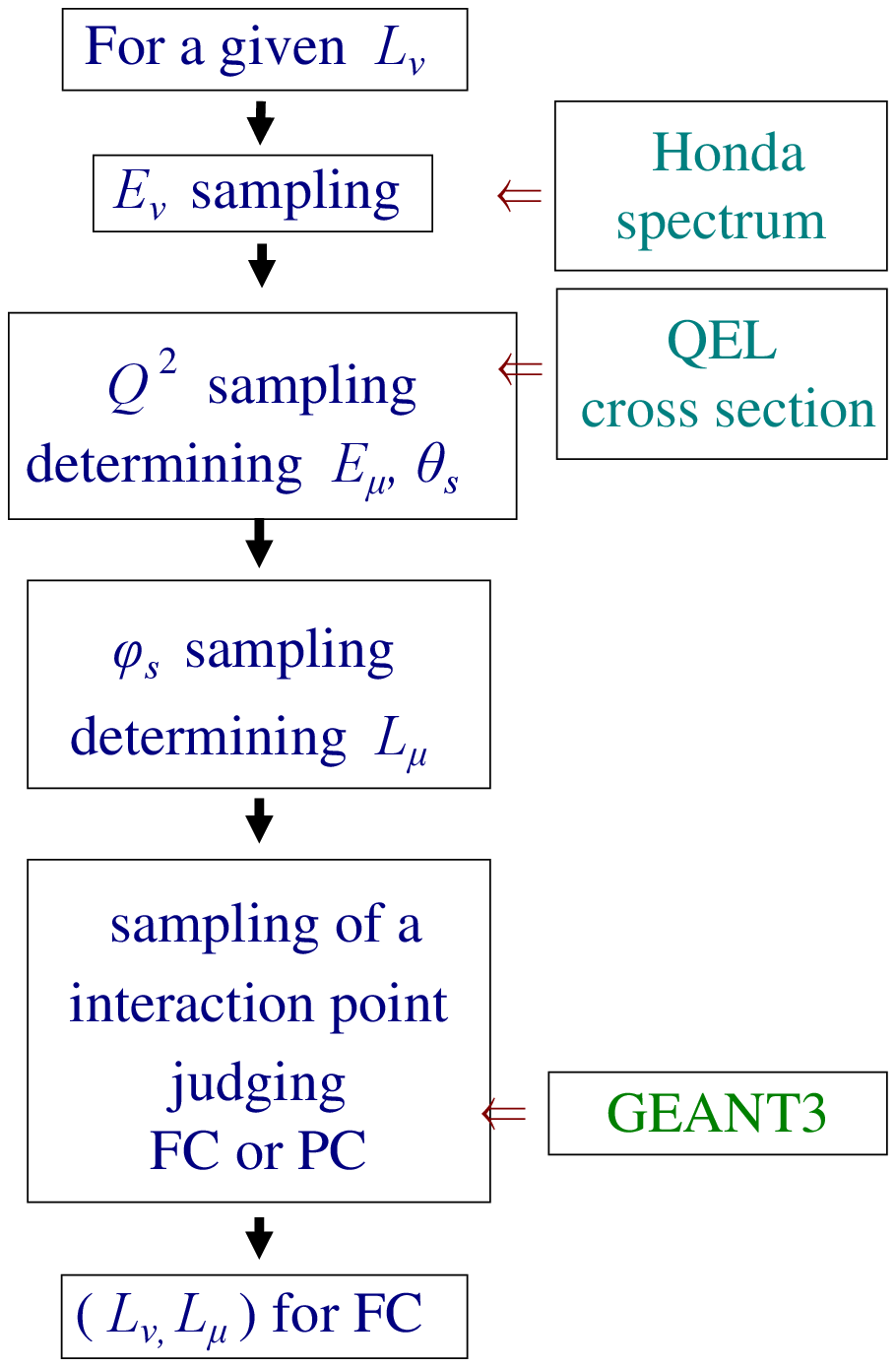}
}
\vspace{-1.5cm}
\caption{The procedure for our numerical experiment
for obtaining $L_{\mu}$ from a given $L_{\nu}$.
}
\label{figR015}
\end{center}
\end{figure}

\subsection
{The correlation between
$L_\nu$  and $L_\mu$
}
 What we need for our argument on the $L/E$ analysis of neutrino 
oscillations are $L_\nu$  and $L_\mu$, 
but not $cos\theta_\nu$ and $cos\theta_\mu$,
 Therefore, we must transform $cos\theta_\nu$ and 
$cos\theta_\mu$ to $L_\nu$  and $L_\mu$. 
 The transformations are carried out by the following equations.
$$
L_{\nu}= R_g \times ( r_{SK} cos\theta_{\nu} +
\sqrt{ r_{SK}^2 cos^2\theta_{\nu} + 1 - r_{SK}^2} )  \,\,\,(7-1)
$$
$$
L_{\mu}= R_g \times ( r_{SK} cos\theta_{\mu} +
\sqrt{ r_{SK}^2 cos^2\theta_{\mu} + 1 - r_{SK}^2} )  \,\,\,(7-2)
$$
\noindent where $R_g$ is the radius of the Earth and 
$r_{SK}=1-D_{SK}/R_g$, $D_{SK}$, 
is the depth of the underground detector from the surface of the 
Earth\cite{Ashie2}.
 In Figure~\ref{figR014}, we give schematically the mutual relation 
among $L_\nu$,  $L_\mu$,  $\theta_s$(the scattering angle due to 
QEL) and $\phi_s$ (the azimuthal angle in QEL). 
As we are exclusively interested in the analysis of {\it Fully 
Contained Events}, we give the procedure for getting the correlation 
between $L_\nu$ and $L_\mu$ for those events in Figure~\ref{figR015}.

\begin{figure}
\begin{center}
\vspace{-1.0cm}
\hspace*{0.5cm}
\resizebox{0.45\textwidth}{!}{%
  \includegraphics{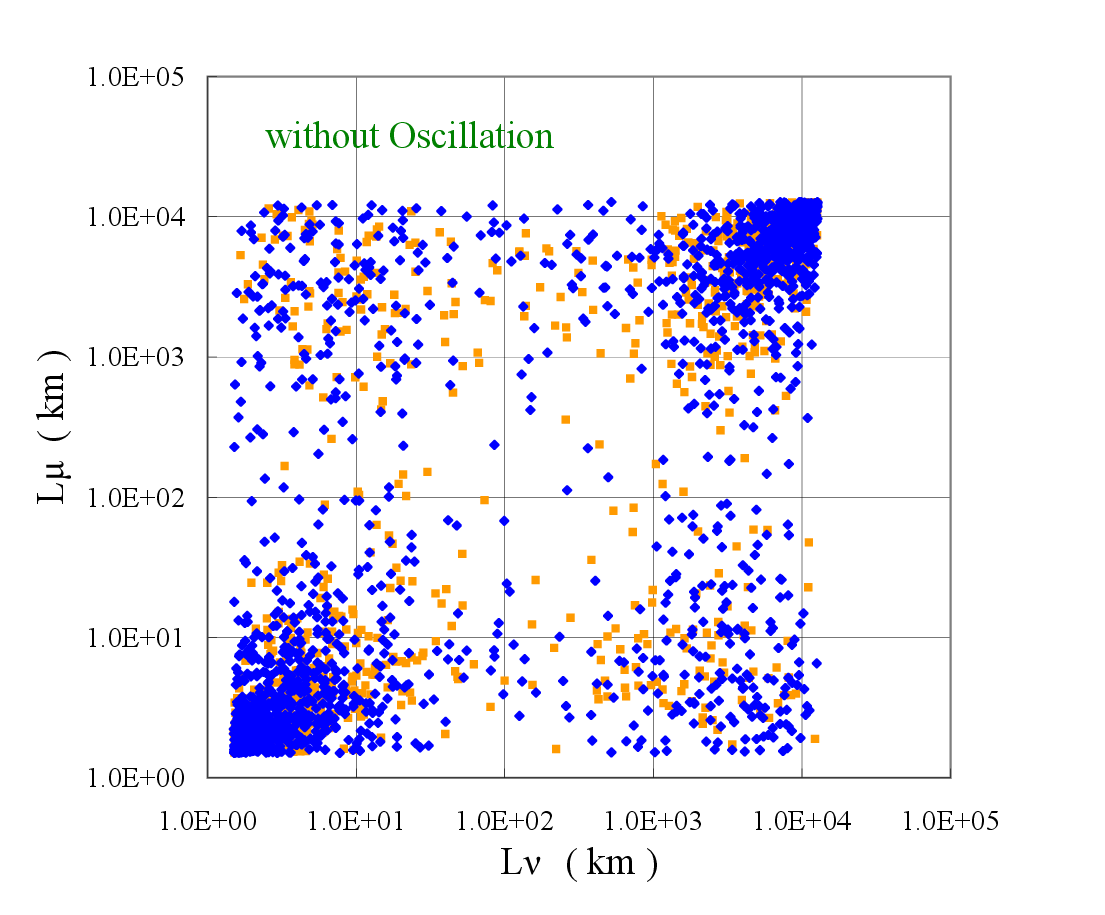}
}
\vspace{-0.8cm}
\caption{The correlation diagram for $L_{\nu}$ and $L_{\mu}$ without 
oscillations for 1489.2 live days of observation.
The blue points and orange points denote neutrino events and 
ani-neutrino events, respectively.  
}
\label{figR016}       
\vspace{0.0cm}
\hspace*{0.5cm}
\resizebox{0.45\textwidth}{!}{%
  \includegraphics{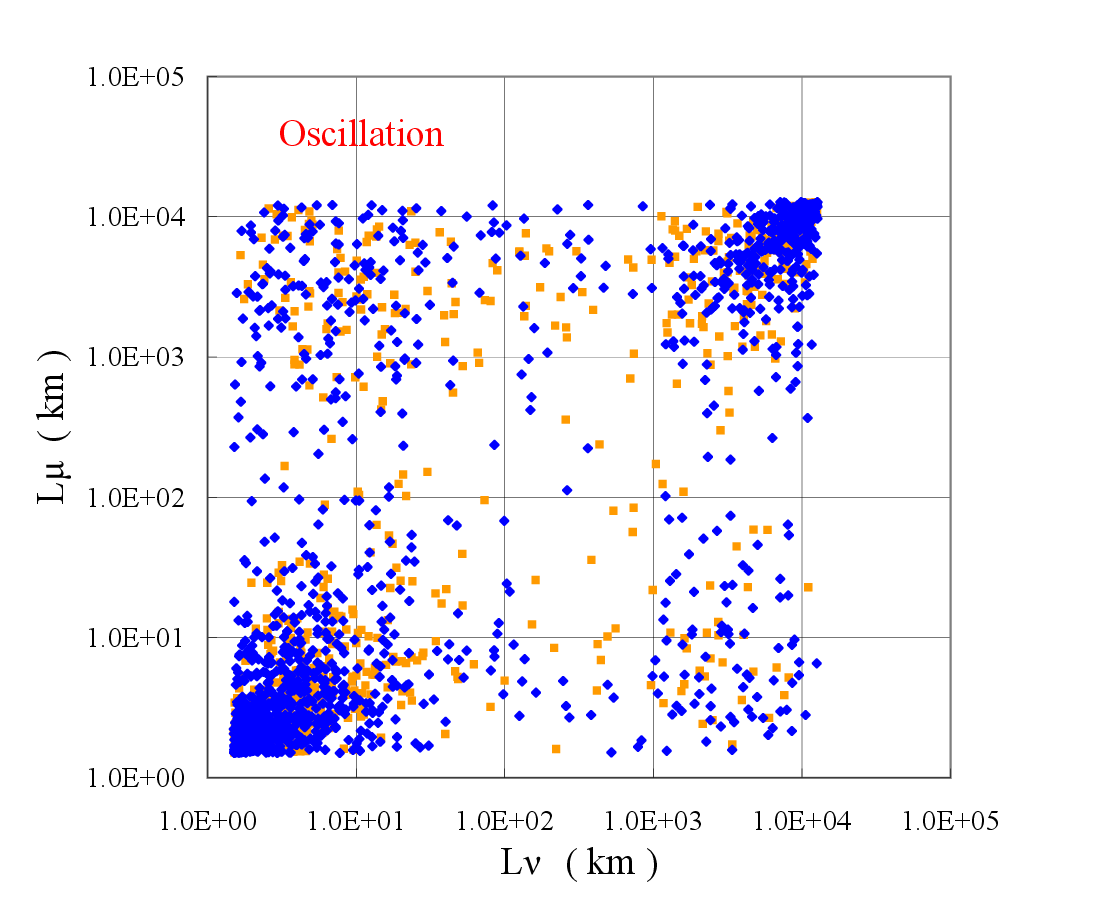}
}
\vspace{-0.8cm}
\caption{The correlation diagram for $L_{\nu}$ and $L_{\mu}$ with 
oscillations for 1489.2 live days of observation.
The blue points and orange points denote neutrino events and 
ani-neutrino events, respectively.
}
\label{figR017}
\vspace{0.0cm}
\hspace*{0.5cm}
\resizebox{0.45\textwidth}{!}{%
  \includegraphics{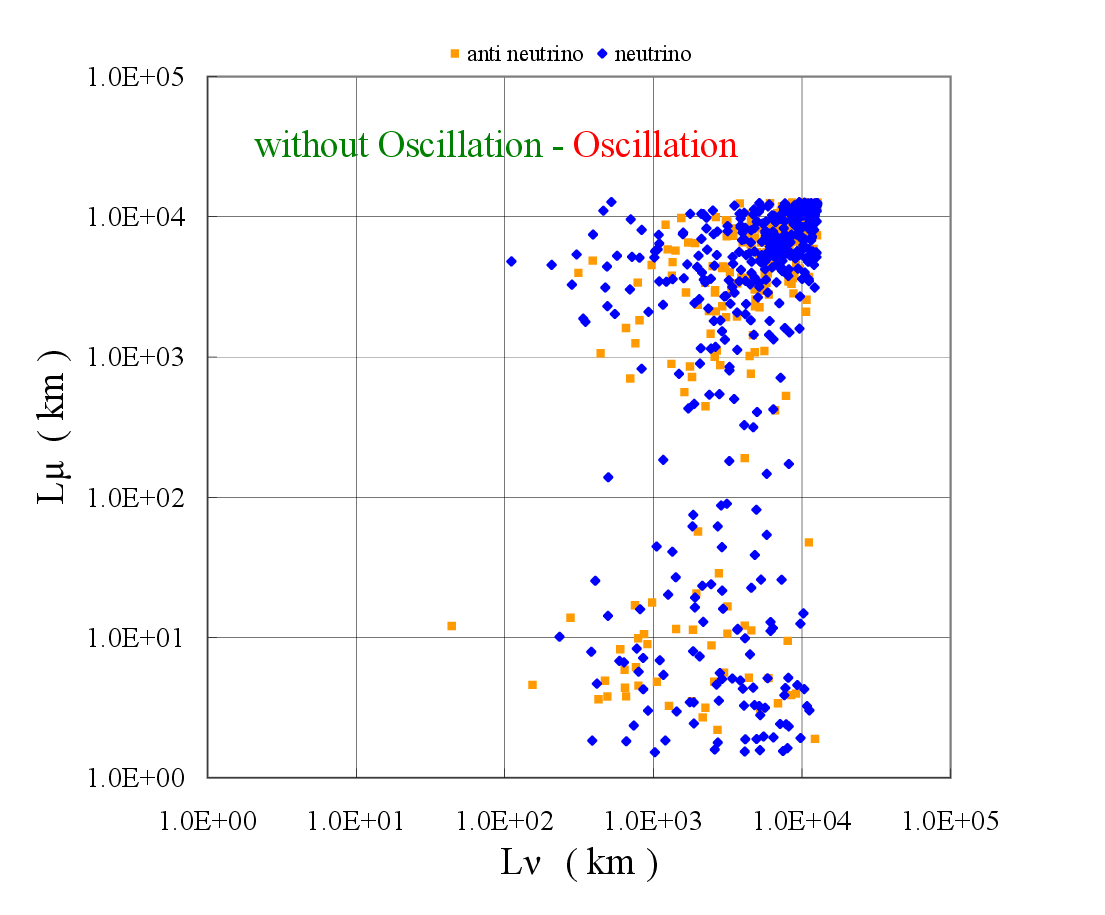}
}
\vspace{-0.8cm}
\caption{The correlation diagram for $L_{\nu}$ and $L_{\mu}$ 
for those events which exist in null oscillations but disappear due to  
oscillations for 1489.2 live days of observation.
The blue points and orange points denote neutrino events and 
ani-neutrino events, respectively.
}
\label{figR018}
\end{center}
\end{figure}

Figures~\ref{figR016}, \ref{figR017} and \ref{figR018} show correlation 
diagrams for $L_{\nu}$ and $L_{\mu}$ obtained by applying the above 
transformation, eqs. (7-1) and (7-2), to the events shown in 
Figures~\ref{figR011}, \ref{figR012} and \ref{figR013}. 

The correlation diagrams expressed by $L_{\nu}$ and $L_{\mu}$ are 
classified into four sectors, just the same as those 
used for the $cos\theta_\nu - cos\theta_\mu$ plot.
Here the coordinate axes, which correspond to both $cos\theta_\nu=0$ 
and $cos\theta_\mu=0$ in Figures~\ref{figR011}, \ref{figR012} 
and \ref{figR013}, 
are $L_{\nu} = L_h$ and $L_{\mu} = L_h$, where 
the numerical value is $L_h \approx 138$~km, 
the distance from the underground detector to the edge of the Earth 
in the horizontal direction.  
We now examine the four sectors in the $L_{\nu} - L_{\mu}$  plot 
(above the horizon or below the horizon), 
as we examined the four sectors in Figures~\ref{figR011}, \ref{figR012} 
and \ref{figR013}, in the $cos\theta_\nu - cos\theta_\mu$ plot 
(downward or upward).  Taking $L_\nu = L_\mu \sim 138$~km 
as the origin of the coordinates, each sector is classified as follows:
\begin{itemize}
\item[(E)] The first sector where both $L_\nu > L_h$ and
$L_\mu > L_h$, respectively. 
This sector corresponds exactly to the first sector 
in Figures~\ref{figR011}, \ref{figR012} 
and \ref{figR013}, in the $cos\theta_\nu - cos\theta_\mu$ plot. 
\item[(F)] The second sector where  $L_\nu < L_h$ and 
$L_\mu > L_h$. 
This sector corresponds exactly to the second sector 
in Figures~\ref{figR011}, \ref{figR012} 
and \ref{figR013}, in the $cos\theta_\nu - cos\theta_\mu$ plot. 
\item[(G)] The third sector where $L_\nu < L_h$ and 
$L_\mu < L_h$.
 This sector corresponds exactly to the third sector 
in Figures~\ref{figR011}, \ref{figR012} 
and \ref{figR013}, in the $cos\theta_\nu - cos\theta_\mu$ plot. 
\item[(H)] The fourth sector where $L_\nu > L_h$ and   
$L_\mu < L_h$.
 This sector corresponds exactly to the fourth sector 
in Figures~\ref{figR011}, \ref{figR012} 
and \ref{figR013}, in the $cos\theta_\nu - cos\theta_\mu$ plot. 
\end{itemize}

\begin{figure*}
\vspace{-1.0cm}
\hspace*{0cm}
\resizebox{0.4\textwidth}{!}{%
  \includegraphics{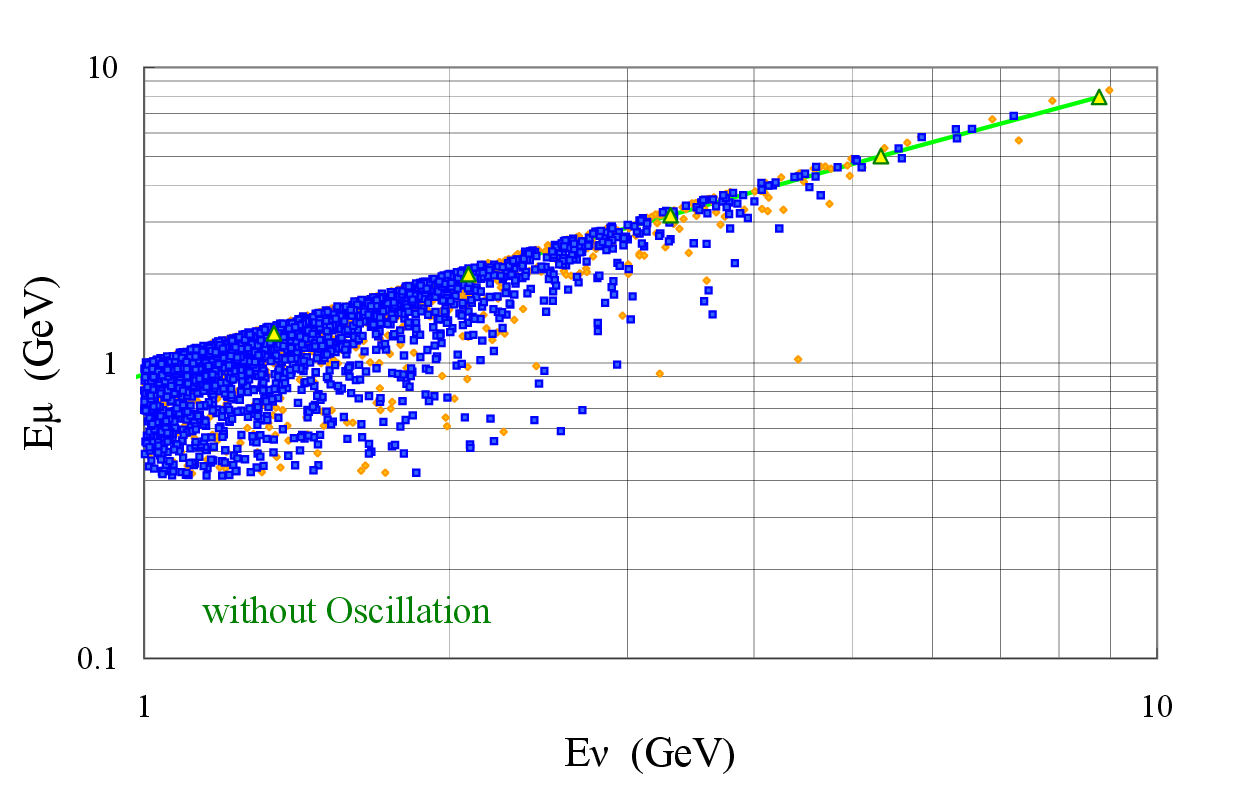}
}
\hspace*{1cm}
\resizebox{0.4\textwidth}{!}{%
  \includegraphics{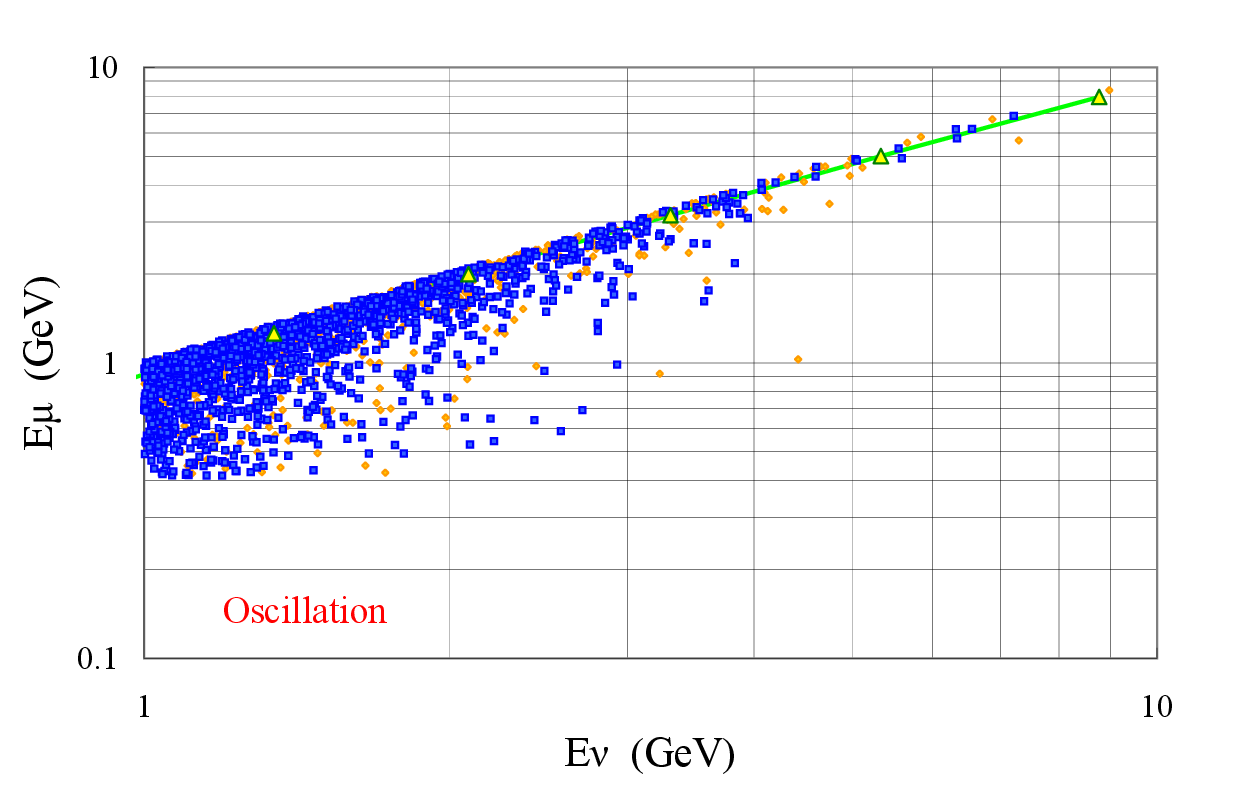}
}
\vspace*{-0.3cm}
\begin{center}
\hspace*{-2cm}(a)
\hspace*{8cm}(b)
\vspace*{-0.0cm} \\
\hspace*{-1cm}
\resizebox{0.4\textwidth}{!}{%
  \includegraphics{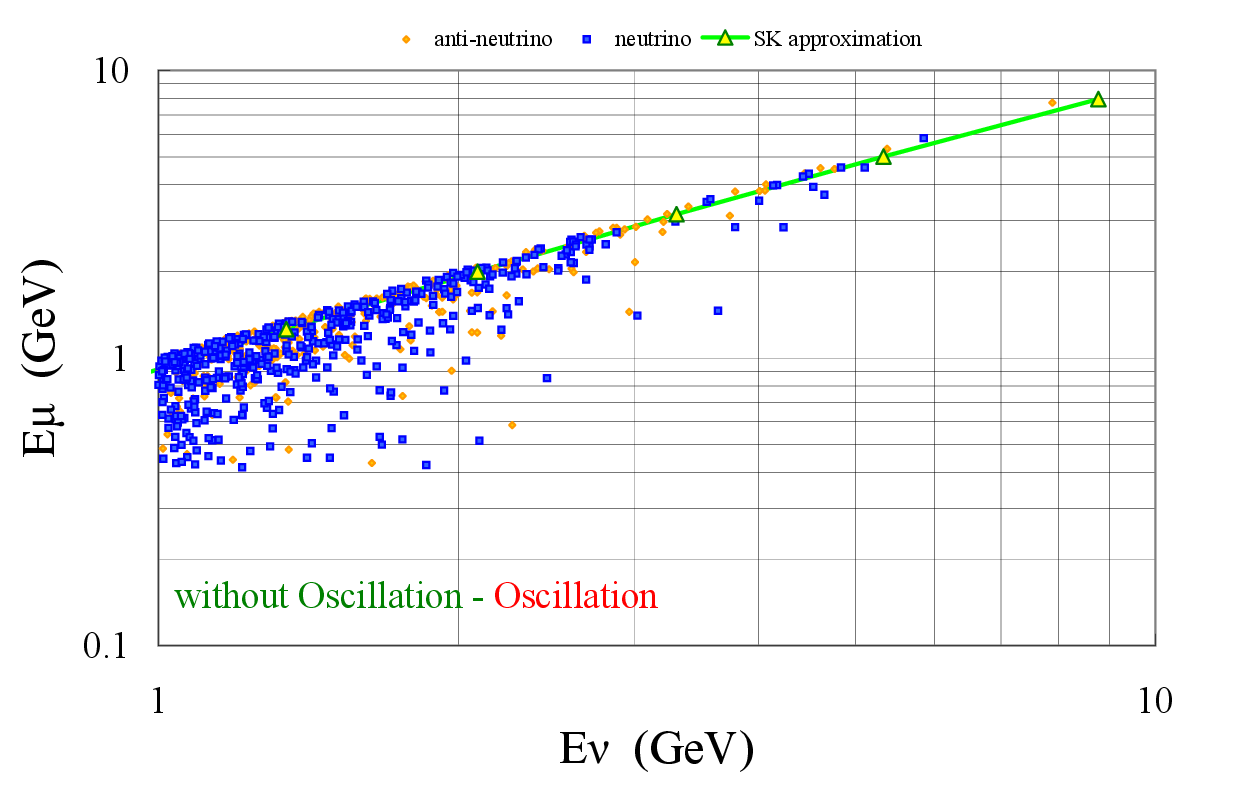}
}
\\
\vspace*{-0.0cm}
(c)
\caption{
The correlation diagram for $E_{\nu}$ and $E_{\mu}$,
(a)without oscillations, (b)with oscillations and 
(c)for those events which exist in null oscillations but 
disappear due to oscillations, for 1489.2 live days of observation, 
respectively. The solid line denotes the 
polynomial expression used by the Super-Kamiokande Collaboration.
}
\label{figR019}
\end{center}
\end{figure*}

\subsubsection
{The case without neutrino oscillations
}
 In Figure~\ref{figR016}, we give the correlation between
$L_{\nu}$ and $L_{\mu}$ in the case without oscillations.
The figure is another expression of Figure~\ref{figR011}. 
The symmetry between the first sector and the third sector, 
and that between the second and the fourth with regard to the origin 
($L_{\nu}=L_{\mu} \sim 138$km) are retained just as in 
Figure~\ref{figR011} with regard to the origin 
($\cos{\theta}_{\nu}=0,\,  \cos{\theta}_{\mu}=0$)
for the same reason.
The density of the number of events around the origin in the 
$L_{\nu}-L_{\mu}$ plot (Figure~\ref{figR016}) 
looks to be far smaller than that in 
$\cos{\theta}_{\nu}-\cos{\theta}_{\mu}$ plot (Figure~\ref{figR011}).
 This is simply due to the fact that the region of 
 $\cos{\theta}_{\nu}$($\cos{\theta}_{\nu}$) $\sim 0$ becomes very wide 
when $L_{\nu}$ ($L_{\mu}$) are expressed on a log-scale 
(see Eq.(7.1) and (7.2)).

\subsubsection
{The case with neutrino oscillations
}
 In Figure~\ref{figR017}, we give the correlation between
$L_{\nu}$ and $L_{\mu}$ in the case without oscillations.
The figure is another expression of Figure~\ref{figR012}.
  The symmetry which exists in Figure~\ref{figR016} is lost here in 
Figure~\ref{figR017} due to the neutrino oscillations just as was the 
case in Figure~\ref{figR011}.  
Also, the situation for the disappeared events which is shown in 
Figure~\ref{figR018} is exactly the same as that shown in 
Figure~\ref{figR013}.
It is easily understood from Figure~\ref{figR018} that the 
disappeared events are found in the region
$L_{\nu}>L_h$ (from under the horizon) 
and they are found mostly in the region
$L_{\mu}>L_h$ (forward scattering) 
while they are scarcely found in the region
$L_{\mu}<L_h$ (backward scattering).
Thus, we reach the important conclusion  from 
Figure~\ref{figR016} to \ref{figR018} that the relation of
$L_{\nu} \approx L_{\mu}$ does not hold,
 irrespective of whether there are oscillations or not,
 just as $\cos{\theta}_{\nu} \ne \cos{\theta}_{\nu}$
shown in Figures~\ref{figR011} to \ref{figR013}.

Here, we return to the question of the validity of Eq.(2). 
The question whether $L_{\nu} \approx L_{\mu}$ or not is classified, 
furthermore, into two sub-questions. One is whether the relation of   
$L_{\nu} \approx L_{\mu}$ holds for each event and the 
other is whether the relation holds only statistically.
 From Figures~\ref{figR016} to \ref{figR018} we see that the above 
relation does not hold even in the statistical sense, 
because $L_{\mu}$ is distributed widely for any given $L_{\nu}$. 
We can conclude that one of the necessary conditions for the validity 
of Eq.(2) in the case A does not hold.

\subsection
{The correlation between
$E_\nu$  and $E_\mu$
}
Now, we examine another necessary condition around $E_\nu$ and $E_\mu$. 
In Figures~\ref{figR019}, we give correlation diagrams of
$E_\nu$ and $E_\mu$ for the events. The solid line in the figures 
is a polynomial equation, which gives the relationship between
$E_\nu$ and $E_\mu$, used by the Super-Kamiokande 
collaboration\cite{Ishitsuka}.
 In Figure~\ref{figR019}-(a) we give the correlation for the events 
without oscillations, while in Figure~\ref{figR019}-(b) we give those for 
the events with oscillations. In Figure~\ref{figR019}-(c), we give 
the correlation for the absent events due to neutrino oscillations.
 It is clear from these figures that the lower 
the energy of the incident neutrinos, the stronger the fluctuations in 
the emitted muon energy, which is easily conjectured from 
Figures~\ref{figH001} and \ref{figH002}, and Table~1.
 Then we can conclude that $E_\nu \approx E_\mu$  does not hold 
irrespective of oscillations or no oscillations, just as in the case of 
the directions.

\subsection
{Summary of section~4
}
There is a distinctive difference in their respective properties 
between $L_\nu$ versus $L_\mu$ and $E_\nu$ versus $E_\mu$  
from the point of view of their correlations.
It is clear from the comparison of Figures~\ref{figR016} to 
\ref{figR018} 16, 17 and 18 with 
Figures~\ref{figR019}-(a), \ref{figR019}-(b) and \ref{figR019}-(c) 
that the difference of muon data from neutrino data is far larger 
in the flight path lengths than in the energies. Consequently, we may 
approximate $E_\nu$ by $E_\mu$ legitimately, but 
it is impossible to approximate $L_\nu$ by $L_\mu$ in any sense.
 The incapability of the replacement of $L_\nu$  
by $L_\mu$ comes from the fact that the azimuthal angle of the 
emitted muons plays an essential role in the transformation
 of $L_\nu$ from $L_\mu$.
 In conclusion, we can definitely say that neither $L_\nu \approx L_\mu$  
nor $E_\nu \approx E_\mu$  can be assumed in the analysis 
of {\it Fully Contained} neutrino events, so that we finally exclude the 
Case A.

\begin{figure}
\begin{center}
\vspace{-1.0cm}
\hspace*{0.5cm}
\resizebox{0.42\textwidth}{!}{%
  \includegraphics{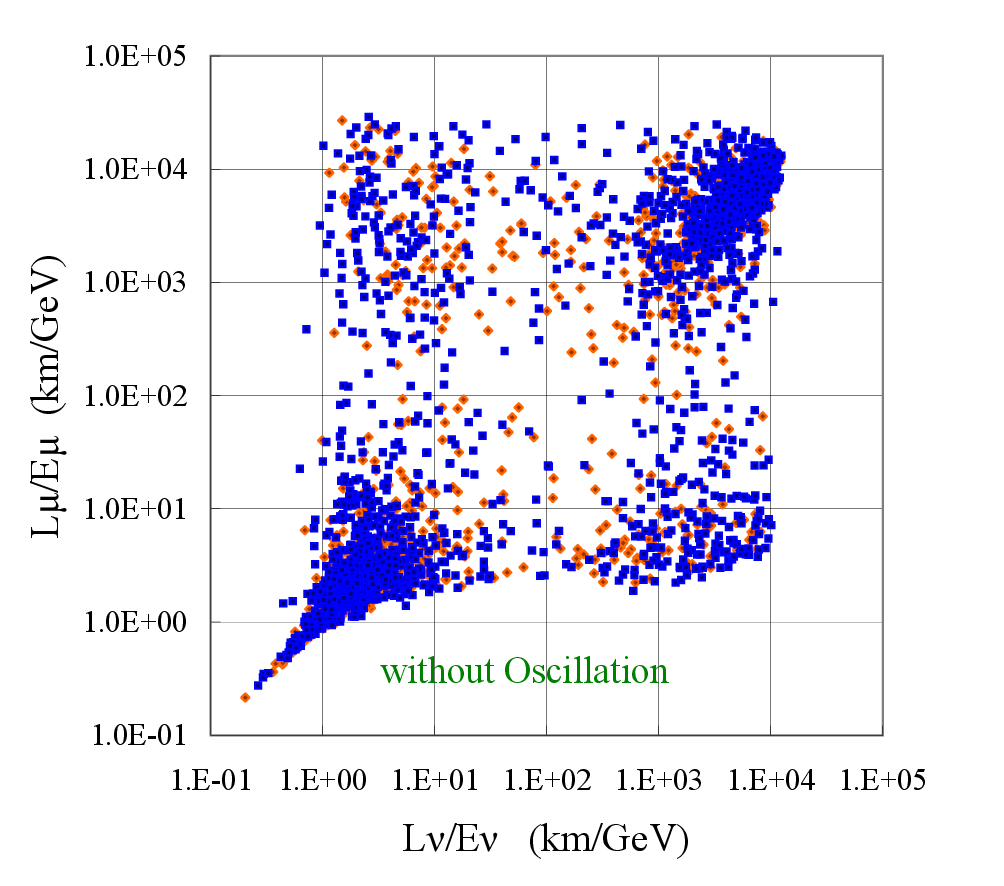}
}
\vspace{-0.3cm}
\caption{The correlation diagram for $L_{\nu}/E_{\nu}$ and 
$L_{\mu}/E_{\mu}$ without 
oscillations for 1489.2 live days of observation.
The blue points and orange points denote neutrino events and 
ani-neutrino events, respectively.  
}
\label{figR020}       
\vspace{0.0cm}
\hspace*{0.5cm}
\resizebox{0.42\textwidth}{!}{%
  \includegraphics{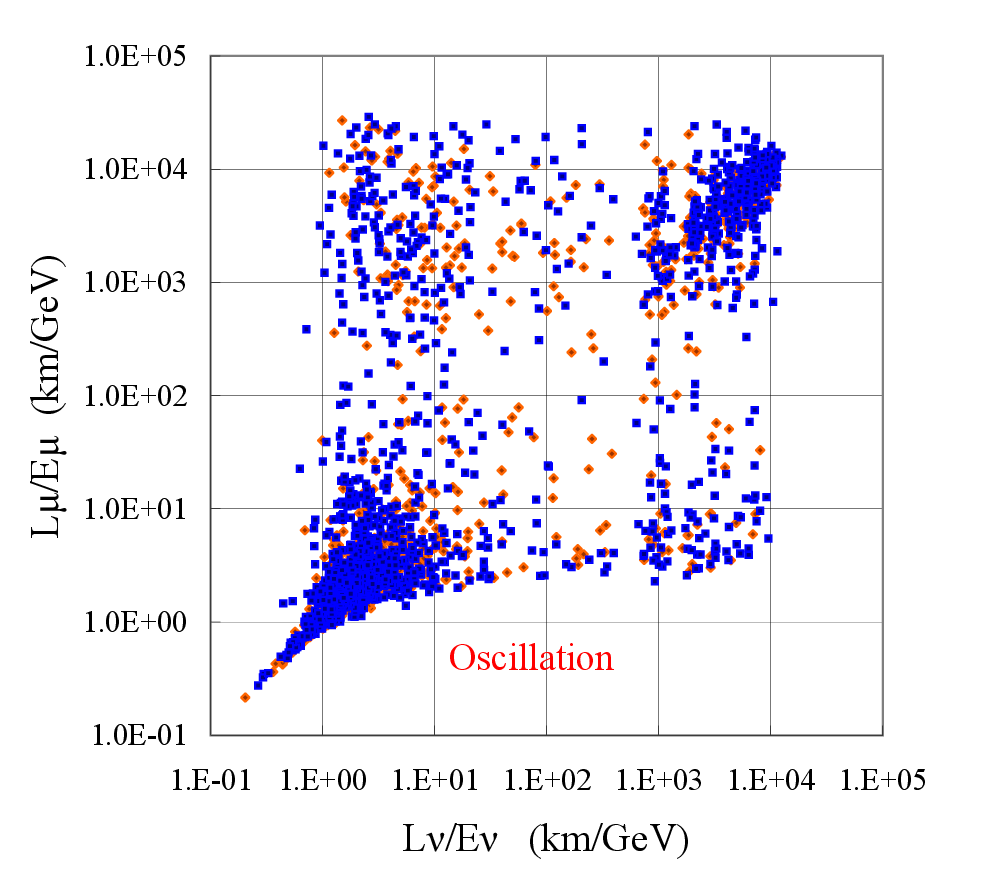}
}

\vspace{-0.3cm}
\caption{The correlation diagram for $L_{\nu}/E_{\nu}$ and 
$L_{\mu}/E_{\mu}$ with 
oscillations for 1489.2 live days of observation.
The blue points and orange points denote neutrino events and 
ani-neutrino events, respectively.
}
\label{figR021}
\vspace{0.0cm}
\hspace*{0.5cm}
\resizebox{0.42\textwidth}{!}{%
  \includegraphics{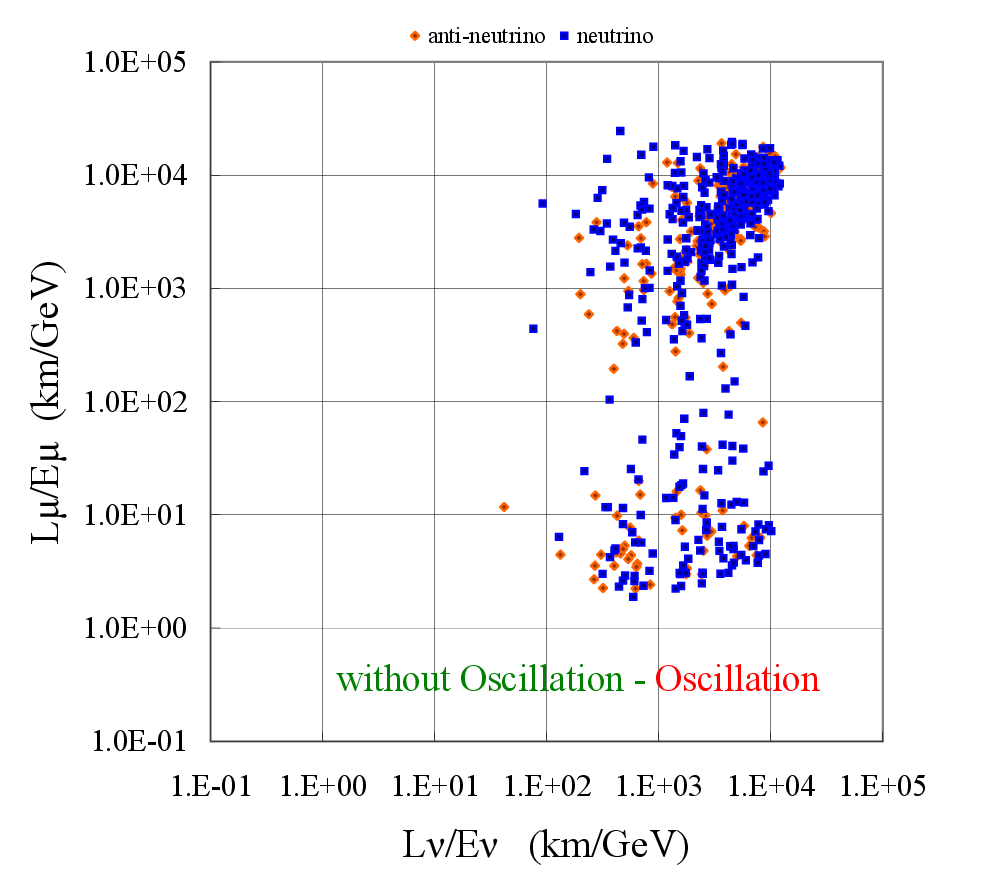}
}
\vspace{-0.3cm}
\caption{The correlation diagram for $L_{\nu}/E_{\nu}$ and 
$L_{\mu}/E_{\mu}$ 
for those events which exist in null oscillations but disappear due to  
oscillations for 1489.2 live days of observation.
The blue points and orange points denote neutrino events and 
ani-neutrino events, respectively.
}
\label{figR022}
\end{center}
\end{figure}

\section
{The correlation between
$L_\nu/E_\nu$ and $L_\mu/E_\mu$
}
In Figures~\ref{figR020}, \ref{figR021} and \ref{figR022}, we give the 
correlation diagram between $L_\nu/E_\nu$ and $L_\mu/E_\mu$ for the events 
without oscillations, with oscillations, and the disappeared events, 
respectively. 

We can give a clear physical image in each section from the first to the fourth with regard to 
the origin ($L_{\nu}=L_{\mu} \sim 138$~km) in the $L_{\nu}-L_{\mu}$
 plot, but we cannot give the corresponding sector in the
$L_\nu/E_\nu$ and $L_\mu/E_\mu$ plot, 
because we cannot define the origin in the plot. 
However, in spite of the clear qualitative difference between the 
$L_\nu/E_\nu-L_\mu/E_\mu$ plot and the $L_{\nu}-L_{\mu}$ plot,
there exist certain similarities between the two 
(compare Figure~\ref{figR020} with Figure~\ref{figR016}, 
Figure~\ref{figR021} with Figure~\ref{figR017}, and 
Figure~\ref{figR022} with Figure~\ref{figR018}). 
Such similarities tell us 
that the characteristics of the correlation between 
$L_\nu/E_\nu$ and $L_\mu/E_\mu$ is mainly determined by 
those between $L_\nu$ and $L_\mu$. 

Here we examine whether the CASE B holds or not in the correlation between $L_\nu/E_\nu$ and $L_\mu/E_\mu$. 
The correlations between $L_\nu/E_\nu$ and $L_\mu/E_\mu$ are more complicated than those between $L_\nu$ and $L_\mu$,
 because they are something like the results of the correlation between   
$L_\nu$ and $L_\mu$ multiplied by the correlation 
between  $E_\nu$ and $E_\mu$.
It is clear from Figures~\ref{figR020} to \ref{figR022} that
$L_\mu/E_\mu$ is distributed widely for any given $L_\nu/E_\nu$, 
irrespective of oscillation or no oscillation and we can conclude that 
the relation $L_\nu/E_\nu \approx L_\mu/E_\mu$
 does not hold in any case.

\section
{Conclusions and Outlook
}
It is clear from the examination in section~4.4 and in section~5 that 
neither CASE A nor CASE B holds.  
Namely, Eq.(2) never holds in any sense. Consequently, if our conclusion 
that Eq.(2) doesn't hold is right, then the only way left for us is 
CASE C. What is CASE C? It may be that we can find neutrino oscillations 
through $L/E$ analysis based on the survival probability in spite of the 
failure of Eq.(2). 
In the present paper, we clarify the failure of Eq.(2) from the analysis 
of the QEL events among the {\it Fully Contained Events}, 
namely, the most reliable events to detect the survival probability itself.
In subsequent papers, we carry out the $L/E$ analysis with the Numerical 
Experiment under the situation that both CASE A and CASE B don't hold, 
including seeking the possibility of CASE C. Also, in subsequent 
papers, we analyze the $L/E$ distribution for all possible combinations of the events, namely, the analyses of 
$L_\nu/E_\nu$, $L_\nu/E_\mu$, $L_\mu/E_\nu$ and $L_\mu/E_\mu$ trying to find the maximum oscillation in the survival probability.

\vspace*{1.0cm}
{\bf Acknowledgment:}
{The authors would like to express their sincere thanks to Prof.M.Tamada 
for his checking and giving suitable comments to the content of their 
paper and to Prof. P.K.Mackeown for his careful reading and 
improving expressions in English of their paper.
}

%

%
%

%
%

%

\newpage
\bibliographystyle{model1a-num-names}

\newpage
\noindent {\bf APPENDICES}\\
\appendix
\section{ 
 Monte Carlo Procedure for the Decision of Emitted Energies of the Leptons and Their Directions }
\setcounter {equation} {0}
\setcounter {figure} {0}
\def\theequation{\Alph{section}\textperiodcentered\arabic{equation}}

Here, we give the Monte Carlo simulation procedure for obtaining the energy and its direction cosines,
$(l_{r},m_{r},n_{r})$, of the emitted lepton in QEL for a given energy and its direction cosines, $(l,m,n)$, of the incident neutrino. 

\begin{figure}
\begin{center}
\resizebox{0.45\textwidth}{!}{%
  \includegraphics{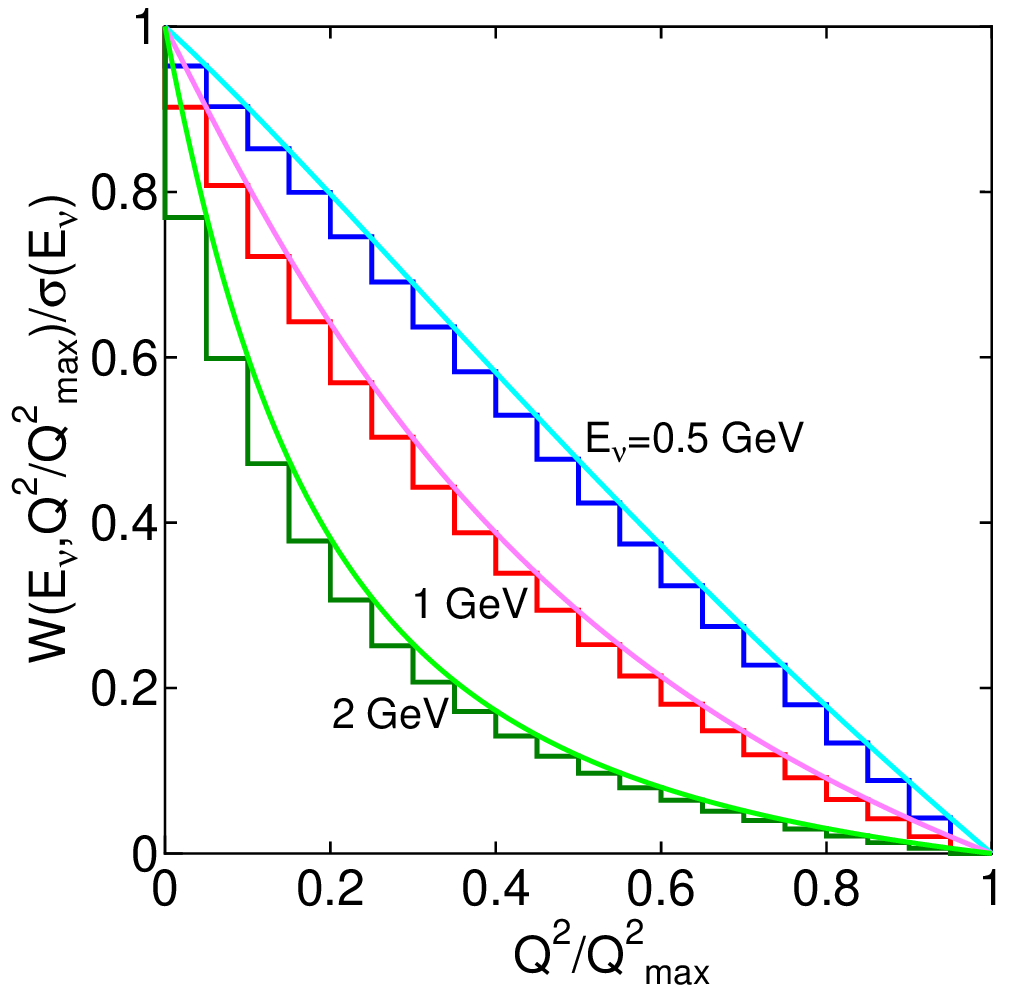}
  }
\end{center}
\caption{The reproduction of the probability function for QEL cross 
section. Histograms are sampling results, while the 
curves concerned are the theoretical ones for given incident energies.
}
\label{figP001}
\end{figure}

The relation among $Q^2$, $E_{\nu(\bar{\nu})}$, the energy of the incident neutrino, $E_{\ell(\bar{\ell})}$, the energy of the emitted lepton (muon or electron or their anti-particles) and $\theta_{\rm s}$, the scattering angle of the emitted lepton, is given as
      \begin{equation}
         Q^2 = 2E_{\nu(\bar{\nu})}E_{\ell(\bar{\ell})}(1-{\rm cos}\theta_{\rm s}).
\label{eqn:a1}  
      \end{equation}
\noindent Also, the energy of the emitted lepton is given by
      \begin{equation}
         E_{\ell(\bar{\ell})} = E_{\nu(\bar{\nu})} - \frac{Q^2}{2M}.
\label{eqn:a2}  
      \end{equation}
\noindent {\bf Procedure 1}\\
\noindent
We select $Q^2$ from the probability function for the differential cross section with a given $E_{\nu(\bar{\nu})}$ (Eq. (\ref{eqn:2}) in the text) by using the uniform random number, ${\xi}$, on (0,1) and solving\\
  \begin{equation}
    \xi = \int_{Q_{\rm min}^2}^{Q^2}P_{\ell(\bar{\ell})}(E_{\nu(\bar{\nu})},Q^2)
                             {\rm d}Q^2,
\label{eqn:a3}  
  \end{equation}
\noindent where
  \begin{eqnarray}
\lefteqn{     P_{\ell(\bar{\ell})}(E_{\nu(\bar{\nu})},Q^2) =} \nonumber \\
&&  \frac{ {\rm d}\sigma_{\ell(\bar{\ell})}(E_{\nu(\bar{\nu})},Q^2) }{{\rm d}Q^2} 
                     \Bigg /\!\!\!\!
      \int_{Q_{\rm min}^2}^{Q_{\rm max}^2} 
      \frac{ {\rm d}\sigma_{\ell(\bar{\ell})}(E_{\nu(\bar{\nu})},Q^2) }{{\rm d}Q^2} 
             {\rm d}Q^2 . \nonumber \\
&&
\label{eqn:a4}  
   \end{eqnarray}
\\
Figure~\ref{figP001} shows a comparison of the distribution of
$Q^2$ sampled by the above procedure, shown by histograms, and 
theoretical one, shown by solid curves. 
The agreement between the sampling data and the theoretical curves is 
excellent, which shows the validity of the utilized procedure in 
Eq.(A$\cdot$3).

\begin{figure}
\begin{center}
\vspace{-1cm}
\resizebox{0.45\textwidth}{!}{%
  \includegraphics{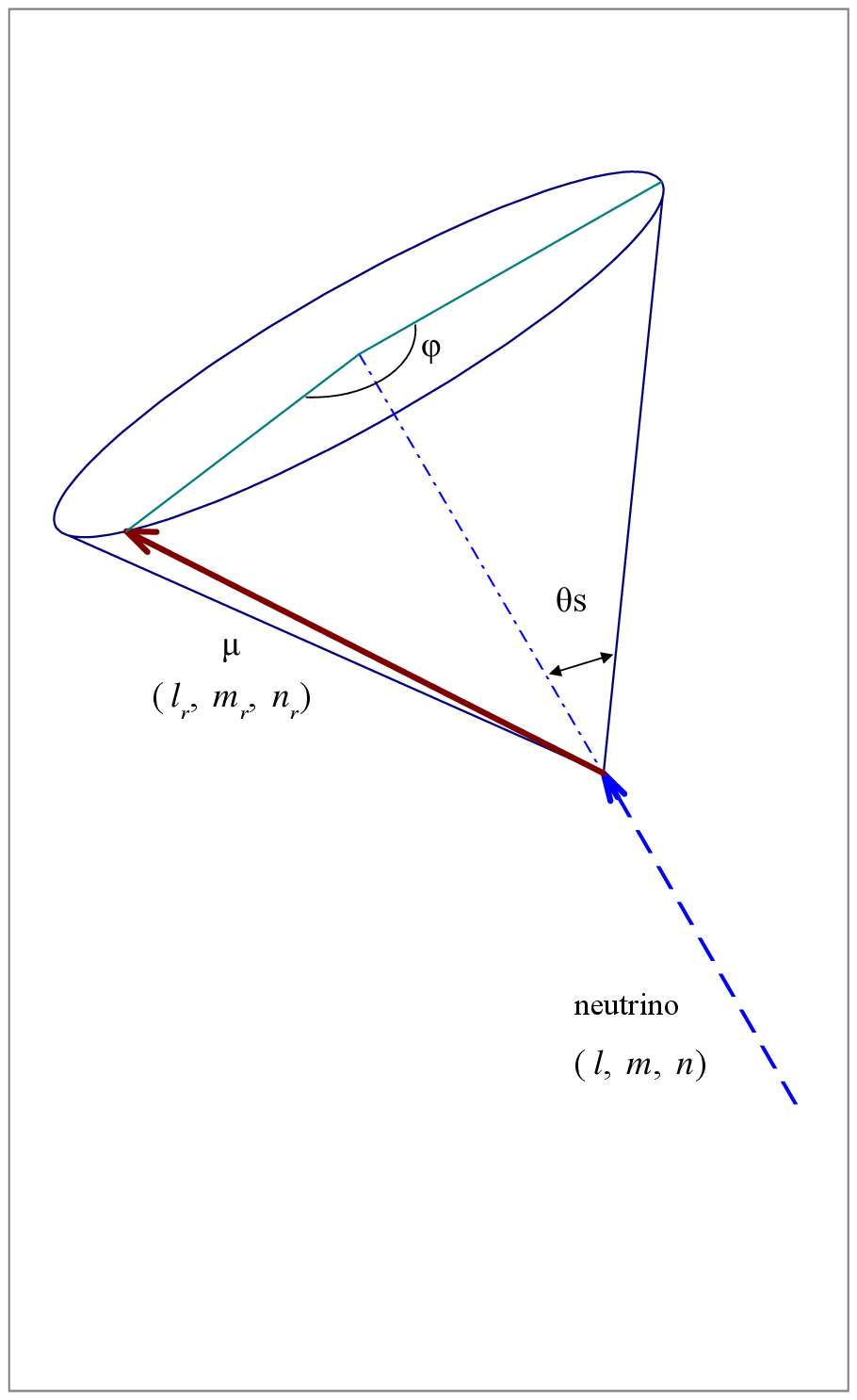}
  }
\end{center}
\caption{ The relation between the direction cosine of the incident neutrino and that of the emitted charged lepton.}

\label{figP002}
\end{figure}

\noindent {\bf Procedure 2}\\
\noindent
We obtain $E_{\ell(\bar{\ell})}$ from Eq. (A$\cdot$2) for  the given $E_{\nu(\bar{\nu})}$ and $Q^2$ obtained as described in Procedure~1.\\

\noindent {\bf Procedure 3}\\
\noindent
We obtain $\cos{\theta_{\rm s}}$, cosine of the the scattering angle of the emitted lepton, for $E_{\ell(\bar{\ell})}$ thus decided in the Procedure 2 from Eq. (A$\cdot$1) in Procedure~2.\\

\noindent {\bf Procedure 4}\\
\noindent
We decide $\phi$, the azimuthal angle of the scattered lepton, which is obtained from\\
  \begin{equation}
       \phi = 2\pi\xi.                 
\label{eqn:a5}  
  \end{equation}
where, $\xi$ is a uniform random number on (0, 1). \\

\noindent {\bf Procedure 5}\\
\noindent
The relation between direction cosines of the incident neutrino, $(\ell_{\nu(\bar{\nu})}, m_{\nu(\bar{\nu})}, n_{\nu(\bar{\nu})} )$, and those of the corresponding emitted lepton, $(\ell_{\rm r}, m_{\rm r}, n_{\rm r})$, for a certain $\theta_{\rm s}$ and $\phi$ is given as
\begin{eqnarray}
\lefteqn{
\left(
         \begin{array}{c}
             \ell_{\rm r} \\
             m_{\rm r} \\
             n_{\rm r}
         \end{array}
       \right)
           =
}
\nonumber\\
    &&   \left(
         \begin{array}{ccc}
           \displaystyle \frac{\ell n}{\sqrt{\ell^2+m^2}} & 
            -\displaystyle 
            \frac{m}{\sqrt{\ell^2+m^2}}        & \ell_{\nu(\bar{\nu})} \\
            \displaystyle \frac{mn}{\sqrt{\ell^2+m^2}} & \displaystyle 
            \frac{\ell}{\sqrt{\ell^2+m^2}}     & m_{\nu(\bar{\nu})}    \\
                        -\sqrt{\ell^2+m^2} & 0 & n_{\nu(\bar{\nu})}
         \end{array}
       \right) \times
\nonumber\\       
 &&      \times
\left(
          \begin{array}{c}
            {\rm sin}\theta_{\rm s}{\rm cos}\phi \\
            {\rm sin}\theta_{\rm s}{\rm sin}\phi \\
            {\rm cos}\theta_{\rm s}
          \end{array}
       \right),
\label{eqn:a6}
\end{eqnarray}

\noindent where $n_{\nu(\bar{\nu})}={\rm cos}\theta_{\nu(\bar{\nu})}$,
 and $n_{\rm r}={\rm cos}\theta_{\ell}$. 
Here, $\theta_{\ell}$ is the zenith angle of the emitted lepton. \\

The Monte Carlo procedure for the determination of $\theta_{\ell}$ of the emitted lepton for the parent (anti-)neutrino with given $\theta_{\nu(\bar{\nu})}$ and $E_{\nu(\bar{\nu})}$ involves the following steps:\\

We obtain $(\ell_r, m_r, n_r)$ by using Eq. (\ref{eqn:a6}). The $n_r$ is the cosine of the zenith angle of the emitted lepton which should be contrasted with $n_{\nu}$, that of the incident neutrino.
\\
Repeating the procedures 1 to 5 just mentioned above, we obtain the zenith angle distribution of the emitted leptons for a given zenth angle of the incident neutrino with a definite energy. \\
\section{ Monte Carlo Procedure to Obtain the Zenith Angle of the Emitted Lepton for a Given Zentith Angle of the Incident Neutrino}

  The present simulation procedure for a given zenith angle of the incident neutrino starts from the atmospheric neutrino spectrum at the opposite site of the Earth to the underground detector. We define,
 $N_{\rm int,no-osc}(E_{\nu(\bar{\nu})},t,{\rm cos}\theta_{\nu(\bar{\nu})})$,
 the interaction neutrino spectrum at the depth $t$ from the 
underground detector for the case no oscillation
in the following way,
   \begin{eqnarray}
    \lefteqn{  N_{\rm int,no_-osc}(E_{\nu(\bar{\nu})},t,
{\cos}\theta_{\nu(\bar{\nu})}) =}\nonumber \\
&&N_{\rm sp}(E_{\nu(\bar{\nu})},
\cos\theta_{\nu(\bar{\nu})}) \times \nonumber \\ 
&&  \Bigg(1-\frac{{\rm d}t}{\lambda_1(E_{\nu(\bar{\nu})},t_1,\rho_1)} \Bigg)  \times\nonumber \\
 &&\times\cdots \times \Bigg(1-\frac{{\rm d}t}{\lambda_n(E_{\nu(\bar{\nu})},t_n,\rho_n)} \Bigg).\nonumber \\
&& 
\label{eqn:b1}
   \end{eqnarray}

Here, $N_{\rm sp}(E_{\nu(\bar{\nu})},\cos\theta_{\nu(\bar{\nu})})$ is the atmospheric (anti-)  neutrino spectrum for the zenith angle at the opposite surface of the Earth.
And $\lambda_i(E_{\nu(\bar{\nu})},t_i,\rho_i)$ 
denotes the mean free path from QEL for the neutrino (anti-neutrino) 
with the energy $E_{\nu(\bar{\nu})}$ at the distance, $t_i$, from the opposite surface of the Earth, where $\rho_i$ is the density. 

In the presence of oscillation, neutrino energy spectrum
correponding to (B$\cdot$1) is given as,

   \begin{eqnarray}
    \lefteqn{  N_{\rm int,osc}(E_{\nu(\bar{\nu})},t,
{\cos}\theta_{\nu(\bar{\nu})}) }\nonumber\\
&&=N_{\rm int,no_-osc}(E_{\nu(\bar{\nu})},
\cos\theta_{\nu(\bar{\nu})}) \times 
P(\nu_{\mu} \rightarrow \nu_{\mu}) 
 \nonumber\\ 
\label{eqn:b2}
   \end{eqnarray}

Here, $P(\nu_{\mu} \rightarrow \nu_{\mu})$ is the survival 
probability of a given flavor, such as $\nu_{\mu}$, and it is given by
                                                    
\begin{eqnarray}
\lefteqn{P(\nu_{\mu} \rightarrow \nu_{\mu})=}
 \nonumber \\
&& 1- sin^2 2\theta \cdot sin^2
(1.27\Delta m^2 L_{\nu} / E_{\nu} ),  
\\ \nonumber 
\end{eqnarray}                                                    

where $sin^2 2\theta = 1.0$ and
$\Delta m^2 = 2.4\times 10^{-3}\rm{eV^2}$ obtained from 
Super-Kamiokande Collaboration\cite{Ashie2}.

The procedures of the Monte Carlo Simulation for the incident neutrino(anti-neutrino) with a given energy, $E_{\nu(\bar{\nu})}$, whose incident direction is expressde by $(l,m,n)$ is as follows.\\

\noindent {\bf Procedure A}\\
\noindent
For the given zenith angle of the incident neutrino, 
${\theta_{\nu(\bar{\nu})}}$, we formulate, 
$N_{\rm pro}( E_{\nu(\bar{\nu})},t,\cos\theta_{\nu(\bar{\nu})}){\rm d}E_{\nu(\bar{\nu})}$,
 the production function for the neutrino flux to produce leptons at the 
Kamioka site as follows.
   \begin{eqnarray}
\lefteqn{N_{\rm pro}( E_{\nu(\bar{\nu})},t,\cos\theta_{\nu(\bar{\nu})}){\rm d}E_{\nu(\bar{\nu})} } \nonumber \\
&&=
 \sigma_{\ell(\bar{\ell})}(E_{\nu(\bar{\nu})}) N_{\rm int}(E_{\nu(\bar{\nu})},t,{\rm cos}\theta_{\nu(\bar{\nu})}){\rm d}E_{\nu(\bar{\nu})},
\nonumber\\
\label{eqn:b2}
   \end{eqnarray}
\noindent where
  \begin{eqnarray}
     \displaystyle
      \sigma_{\ell(\bar{\ell})}(E_{\nu(\bar{\nu})}) = \int^{Q_{\rm max}^2}_{Q_{\rm min}^2}  \frac{ {\rm d}\sigma_{\ell(\bar{\ell})}(E_{\nu(\bar{\nu})},Q^2)}{{\rm d}Q^2}{\rm d}Q^2.
\label{eqn:b3}
\nonumber\\
  \end{eqnarray}

\noindent \\
Each differential cross section above is given by Eq. (\ref{eqn:2}) in the text.
Here, we simply denote the interaction energy spectrum as
$ N_{\rm int}(E_{\nu(\bar{\nu})},t,
{\cos}\theta_{\nu(\bar{\nu})}) $
, irrespective of the absence or the presence of 
oscillation.

Utilizing, $\xi$, a uniform random number on (0,1), 
we determine $E_{\nu(\bar{\nu})}$, the energy of the incident neutrino 
by the following sampling procedure,\\
    \begin{eqnarray}
       \xi = \int_{E_{\nu(\bar{\nu}),{\rm min}}}^{E_{\nu(\bar{\nu})}}
             P_d(E_{\nu(\bar{\nu})},t,\cos\theta_{\nu(\bar{\nu})}(\bar{\nu})){\rm d}E_{\nu(\bar{\nu})}
\nonumber\\
    \end{eqnarray}
where
\begin{eqnarray}
\lefteqn{
        P_d(E_{\nu(\bar{\nu})},t,\cos{\theta}_{\nu(\bar{\nu})}){\rm d}E_{\nu(\bar{\nu})} 
        } \nonumber\\
&=& 
\frac{
N_{pro}( E_{\nu(\bar{\nu})},t,\cos{\theta}_{\nu(\bar{\nu})}){\rm d}E_{\nu(\bar{\nu})} 
}
{ \displaystyle \int_{E_{\nu(\bar{\nu}),{\rm min}}}^{E_{\nu(\bar{\nu}),{\rm max}}} 
                       N_{pro}( E_{\nu(\bar{\nu})},t,\cos{\theta}_{\nu(\bar{\nu})}){\rm d}E_{\nu(\bar{\nu})} 
} .
\nonumber\\
\end{eqnarray}
\\
In our Monte Carlo procedure, the reproduction of, 
$P_d(E_{\nu(\bar{\nu})},t,\cos\theta_{\nu(\bar{\nu})}){\rm d}E_{\nu(\bar{\nu})}$, 
the normalized differential neutrino interaction probability function, is confirmed in the same way as in Eq. (A$\cdot$4). 
\\
 
\noindent {\bf Procedure B}\\
\noindent
For the (anti-)neutrino concerned with the energy of $E_{\nu(\bar{\nu})}$, 
we sample $Q^2$ utlizing a uniform random number $\xi$ on (0,1). 
The Procedure~B is exactly the same as Procedure~1 in Appendix~A. \\

\noindent {\bf Procedure C}\\
\noindent
We select ${\theta_{\rm s}}$, the scattering angle of the emitted lepton 
for given $E_{\nu(\bar{\nu})}$ and $Q^2$. Procedure C is exactly the same 
as in the combination of Procedures~2 and 3 in Appendix~A. \\

\noindent {\bf Procedure D}\\
\noindent
We randomly sample the azimuthal angle of the charged lepton concerned. 
The Procedure~D is exactly the same as in Procedure~4 in Appendix~A. \\
 
\noindent {\bf Procedure E}\\
\noindent
We decide the direction cosine of the charged lepton concerned.  
The Procedure E is exactly the same as Procedure~5 in Appendix~A.\\

We repeat Procedures A to E until we reach the desired number of samples. \\

\section{Correlation between the Zenith Angles of the Incident Neutrinos and Those of the Emitted Leptons}

\noindent {\bf Procedure A}\\
By using, $N_{\rm pro}( E_{\nu(\bar{\nu})},t,\cos\theta_{\nu(\bar{\nu})}){\rm d}E_{\nu(\bar{\nu})}$,
which is defined in Eq. (\ref{eqn:b2}), 
we define the spectrum for $\cos\theta_{\nu(\bar{\nu})}$  in the following.

\begin{eqnarray}
\lefteqn{      I(\cos\theta_{\nu(\bar{\nu})}){\rm d}(\cos\theta_{\nu(\bar{\nu})}) = } \nonumber \\
&&{\rm d}(\cos\theta_{\nu(\bar{\nu})})\times
\nonumber \\
&&   \times
    \int_{E_{\nu(\bar{\nu}),{\rm min}}}^{E_{\nu(\bar{\nu}),{\rm max}}}
 N_{\rm pro}( E_{\nu(\bar{\nu})},t,\cos\theta_{\nu(\bar{\nu})}){\rm d}E_{\nu(\bar{\nu})}.
 \nonumber \\ 
\label{eqn:c1}
\end{eqnarray}

\noindent using Eq.(\ref{eqn:c2}) where $\xi$ is a sampled uniform random 
number on (0,1), we can determine $\cos\theta_{\nu(\bar{\nu})}$
    \begin{equation}
      \xi = \int_0^{\cos\theta_{\nu(\bar{\nu})}}P_n(\cos\theta_{\nu(\bar{\nu})})
                          {\rm d}(\cos\theta_{\nu(\bar{\nu})}),
\label{eqn:c2}
    \end{equation}

\noindent where

\begin{eqnarray}
\lefteqn{
P_n(\cos\theta_{\nu(\bar{\nu})}) =  
}
 \nonumber \\
&&
\frac{   I(\cos\theta_{\nu(\bar{\nu})})
} 
{\displaystyle
 \int_0^1 I(\cos\theta_{\nu(\bar{\nu})}){\rm d}(\cos\theta_{\nu(\bar{\nu})})
}.
\nonumber \\
\label{eqn:c3}
    \end{eqnarray}

\noindent {\bf Procedure B}\\
\noindent
For the sampled ${\rm d}(\cos\theta_{\nu(\bar{\nu})})$ in Procedure A,
 we sample 
$E_{\nu(\bar{\nu})}$ from Eq.(\ref{eqn:c4}) by using ${\xi}$, a uniform randum number on (0,1), 

 \begin{equation}
    \displaystyle
       \xi = \int_{E_{\nu(\bar{\nu}),{\rm min}}}^{E_{\nu(\bar{\nu})}} 
                    P_{pro}(E_{\nu(\bar{\nu})},\cos\theta_{\nu(\bar{\nu})}){\rm d}E_{\nu(\bar{\nu})}, 
\label{eqn:c4}
    \end{equation}
where
      \begin{eqnarray}
\lefteqn{
         P_{pro}(E_{\nu(\bar{\nu})},t,\cos\theta_{\nu(\bar{\nu})}){\rm d}E_{\nu(\bar{\nu})} =
} \nonumber \\
&&  
\frac{         N_{\rm pro}( E_{\nu(\bar{\nu})},t,\cos\theta_{\nu(\bar{\nu})}){\rm d}E_{\nu(\bar{\nu})}
}
{\displaystyle         \int_{E_{\nu(\bar{\nu}),{\rm min}}}^{E_{\nu(\bar{\nu}),{\rm max}}} 
         N_{\rm pro}( E_{\nu(\bar{\nu})},t,\cos\theta_{\nu(\bar{\nu})}){\rm d}E_{\nu(\bar{\nu})}
}.
\nonumber\\
\label{eqn:c5}
      \end{eqnarray}

\noindent\\
\noindent {\bf Procedure C}\\
\noindent 
For the sampled $E_{\nu(\bar{\nu})}$ in Procedure B, we sample $E_{\mu(\bar{\mu})}$ from Eqs. (\ref{eqn:a2}) and  (\ref{eqn:a3}). 
From the sampled  $E_{\nu(\bar{\nu})}$ 
and $E_{\mu(\bar{\mu})}$, we determine $\cos{\theta}_s$, the scattering angle of the muon uniquely from Eq. (\ref{eqn:a1}).\\

\noindent {\bf Procedure D}\\
\noindent
We determine, $\phi$, the azimuthal angle of the scattering lepton from Eq. (\ref{eqn:a5}) by using a uniform randum number ${\xi}$ on (0,1). \\

\noindent {\bf Procedure E}\\
\noindent  
We obtain $\cos{\theta}_{\mu(\bar{\mu})}$ from Eq. (\ref{eqn:a6}).
As the result, we obtain a pair of ($\cos\theta_{\nu(\bar{\nu})}$, 
$\cos{\theta}_{\mu(\bar{\mu})}$) through Procedures A to E. Repeating the 
Procedures A to E, we finally obtain the correlation between the zenith 
angle of the incident neutrino and that of the emitted muon.

\label{}

\end{document}